\documentclass{aa}  
\usepackage{graphicx,rotating}
\usepackage[varg]{txfonts}

\begin{document} 

   \title{The structure of $\kappa$ Cygnid and August Draconid meteoroid 
   streams}%\thanks{The used meteor data are available at the CDS via ...}}

   \author{J.~Borovi\v{c}ka \inst{1} \and P.~Spurn\'y\inst{1} \and  L.~Kotkov\'a \inst{1}\and S.~Molau\inst{2} 
   \and D.~Tomko \inst{3} \and T.~Weiland \inst{4}
          }

   \institute{Astronomical Institute of the Czech Academy of Sciences, Fri\v{c}ova 298, 25165 Ond\v{r}ejov, Czech Republic \\
              \email{jiri.borovicka@asu.cas.cz}
   \and
   Arbeitskreis Meteore e.V., Abenstalstra\ss e 13b, 84072 Seysdorf, Germany
    \and          
           Astronomical Institute of the Slovak Academy of Sciences, 05960 Tatranská Lomnica, Slovak Republic
    \and
         Österreichischer Astronomischer Verein, Laverangasse 40/5, 1130 Wien, Austria
      }

   \date{Received 20 December 2024, Accepted 31 January 2025}
   \titlerunning{$\kappa$ Cygnid and August Draconid meteoroid streams}
   \authorrunning{Borovi\v{c}ka et al.}

\abstract{Meteoroid streams can be complex structures shaped by the processes of their formation and
subsequent orbital evolution. The first step of their understanding is mapping their current stage. 
We used precise data from the European Fireball Network to disentangle the situation with meteor showers
active in August and having radiants in the Cygnus-Draco area. In total, 179 fireballs observed between 2016--2024 were analyzed.
We confirmed that two showers, $\kappa$ Cygnids and August Draconids, are present.
The meteoroid swarm producing $\kappa$ Cygnids is locked in the 5:3 main-motion resonance with Jupiter with
orbital period 7.12 years and has a limited extent of $\leq90\degr$ in mean anomaly. The shower is therefore
markedly active only once or twice during each seven-year period. The orbits have wide range of inclinations, 28--44\degr.
There is a correlation between inclination, perihelion distance, and argument of perihelion due to observational selection effects. 
The radiant area is almost
30\degr\ long in declination. August Draconids have even more extended radiant and can be divided into three branches 
depending on the position of the perihelion relative to the ecliptic plane. Neither of the showers can be described by a single set
of orbital elements. We provide sets of representative orbits and identifications with showers previously reported in the
literature. Physical properties of meteoroids and possible parent bodies are also discussed.
 }

   \keywords{Meteorites, meteors, meteoroids}

   \maketitle

\section{Introduction}

Meteoroid streams are formed by meteoroids on similar orbits originating in the same parent body, typically a comet.
Young streams are compact and narrow and can be easily linked with their parent bodies on the basis
of orbital similarity in most cases. After several thousands years, the streams become dispersed and their mean orbit can differ
from the orbit of the parent body, especially if the parent body or a part of the stream were subject to close encounter with 
Jupiter or another major planet \citep{Vaubaillon}. Meteoroid streams persist even if their parent comet completely disintegrated and remain then
the only witness of comet existence.

When crossing Earth's orbit, meteoroid streams produce meteor showers. Meteor showers are of popular interest and a traditional
subject of meteor astronomy. Their studies provide information about the properties and history of their parent bodies and about
the processes in the inner Solar System. Each meteor shower has its own unique history. The published information
about meteor showers is compiled by the Meteor Data Center of the International Astronomical Union (IAU MDC) \citep{MDC}.
The center maintains the list of established showers, whose existence has been confirmed (currently 110 showers in total),
and the working list containing proposed showers. 
Regrettably, the parameters of many established showers remain poorly known since the 
radiants and orbits published by different authors differ substantially.
Although part of these differences may be real, for example because various techniques are sensitive to meteoroids of various sizes or
because the showers evolve in time, most of them are due to limitations of data samples and observational uncertainties.
A discussion about cross-correlation of meteor showers can be found in \citet{Jopek2024}.

Here we investigate the established shower $\kappa$ Cygnids, listed in the IAU MDC as number 12 with the acronym KCG. 
The activity of the shower lasts for most of August and the shower is known to produce bright fireballs \citep{JennBook1}.
Many authors noted that other showers with radiants nearby are active at the same time and the situation is
therefore complex (see Sect.~\ref{history}). Other proposed showers include the August Draconids, which are now listed as established
shower 197 AUD. We use the recent precise data from the European Fireball Network to study these showers independently
on previous analyses.

The overview of past observations of  $\kappa$ Cygnids and the related showers is provided in Sect.~\ref{history}.
Our goals and data procedures are explained in Sect.~\ref{methods}. The groups of radiants we identified are
defined in Sect.~\ref{groups} and the properties of these potential showers are described in 
Sects.~\ref{understanding}--\ref{periodicity}. The cross-identification of our groups with previously reported showers
is given in Sect.~\ref{identification}. Sect.~\ref{refinement} provides the refined parameters of the showers based on our data.
Physical properties of the meteoroids are evaluated in Sect.~\ref{physical}. The article is then finished by discussion and conclusions.

\section{History of $\kappa$ Cygnid observations}
\label{history}

The recognition of $\kappa$ Cygnid meteor shower is connected with visual observations in Great Britain.
The first strong manifestation was reported by \citet{Denning1879a,Denning1879b} who observed an extraordinary display
from a radiant located at (RA, Del) = (291\degr, +60\degr) on August 21--23, 1879. He noted that the shower was
not new. Meteors were observed from similar radiants in some earlier years too, though probably with lower
frequency than in 1879. Another remarkable display was observed by the same author and other observers in 1893
\citep{Denning1893}. Denning did not connect the two displays. In his catalog of meteor shower radiants 
\citep{Denning1899} he treated these two outbursts as separate showers. The 1879 shower was called
$o$ Draconids with the radiant at (291\degr, +60\degr) and activity August 21--25, while the 1893
shower was called $\theta$ Cygnids with the radiant at (292\degr, +53\degr) and activity on August 4--16.
Later he noted the activity of $o$ Draconids on August 15, 1907 \citet{Denning1907}.

Another similar shower with remarkable activity and a radiant near $\alpha$ Lyrae (280\degr, +44\degr) 
was reported by several observers on August 8--23, 1914 \citep{Davidson1914}. 
The $\theta$ Cygnids were active at that time as well.
The connection between the two showers was discussed, noting similar orbital elements. High activity of 
$\theta$ Cygnids, including fireballs with bright terminal flashes, was further reported on August 15-25, 1922 
\citep{Cook1924}. In that report, high activity on August 8--12, 1901, was also mentioned, though the original report
by \citet{Besley1901} just listed the shower among those observed in August 1901. \citet{Cook1924} 
quoted Mr. Denning considering the shower to have a period of about 7.5 years, probably the first such a note in the literature. 
Fireball presence was also mentioned in 1929 but the overall activity was not very high \citep{Obs1929}.

The Harvard photographic program provided the first five instrumentally determined orbits of the members
of the shower, which was in the meantime renamed to $\kappa$ Cygnids \citep{Whipple1954}. The orbits
were obtained between August 9--22, 1950,
during another year of high activity. The obtained semimajor axes correspond to orbital periods 6.6 -- 10 years. 
\citet{Letfus} computed the orbits of four of these meteors independently and obtained periods 5.5 -- 9 years.

Since the work of \citet{Whipple1954}, $\kappa$~Cygnids have been considered as a confirmed shower but
no special attention was devoted to them in the following four decades. They were listed in the catalog of meteor
showers by \citet{Cook} with geocentric radiant at (286\degr, +59\degr) and orbital period 5.3 years. They were also
detected among faint radar meteors by \citet{Sekanina1973} with an orbital period of 4.2 years only.

A computer search among 3518 published photographic meteor orbits by \citet{Lindblad1995} identified 37 ``Cygnid''
meteors, which could be split into four separate showers when the search was refined. One of them was $\kappa$ Cygnids
which contained nine meteors. Five of them were observed in 1950, and one in 1957, 1958, 1978, and 1993 each.
The dates span August 12--22. 
The mean radiant was at (286\degr, +55\degr) and the mean orbital period was 7.2 years with the range from 5.3 to 10 years.
The other three similar showers were called $\alpha$ Lyrids, $\zeta$ Draconids, and August Lyrids by \citet{Lindblad1995}. 

In between, another outburst was observed visually and photographically by the members of the Dutch Meteor Society (DMS) 
in 1993 \citep{JennBook1}.
\citet{JennBook2} mentioned other outbursts observed by the DMS visually in 1978 and 1985.

\citet{Jenn1994}, compiling visual observations by two groups in 1981--1991,
listed $\kappa$ Cygnids among annual showers with a $\sim 30$ day activity
peaking at solar longitude $\lambda_\sun = 146\degr$ ($\sim$ August 19) with the zenith hourly rate (ZHR) of 2.3.

Similarly to \citet{Lindblad1995}, \citet{Jones2006} performed a computer search of $\kappa$ Cygnid related meteors
among now more extensive meteor catalogs. They confirmed the four Lindblad's substreams of what they called a `Kappa Cygnid meteoroid complex'
and added another one named $\gamma$ Draconids. They also noted that there are no gaps between the locations of radiants of
adjacent substreams. When the IAU MDC, following \citet{JennBook1}, prepared a numbered list of meteor showers with three-letter
acronyms in 2007--2009, only $\kappa$ Cygnids were included (as 12/KCG) from these five showers. The name $\zeta$ Draconids
was reused for another shower (73/ZDR) proposed by \citet{Molau2009} to be active at the end of July.

Another outburst of bright $\kappa$ Cygnids was observed in 2007 \citep{CBET2007}. \citet{Trigo2009} published 
the radiants and orbits of nine meteors. They also noted low tensile strength of the meteoroids, normal chondritic
composition derived from one spectrum, and the absence of faint meteors (magnitude $>$ +4) in the shower.
Nevertheless, visual observations of one of us (TW) in eight nights indicated that
about 20\% of $\kappa$ Cygnids were of magnitude +4 or fainter.

\citet{SonotaCo} published a meteor shower catalog based on 240,000 single-station video meteors observed in 2007--2008.
$\kappa$ Cygnids are listed with the largest radiant area of all showers, having a radius of 10\degr.

\citet{Koseki2014}, using video data from the SonotaCo's network, pointed out high activity of $\kappa$ Cygnids in 2007
 in comparison with years 2008--2012. He also compiled literature data and introduced the term `Cygnids-Draconids Complex' (C-D Complex). 
He divided video radiants within the complex into seven groups, A--G. Group A was in fact active at the end of July and the orbits had
much larger eccentricity than the other groups. It could be identified with a distinct shower, the July $\gamma$ Draconids (184/GDR).
Groups B and C were those which exhibited the high activity in 2007. The author was not sure if this activity was the same shower 
as the activity observed photographically in 1950 \citep{Whipple1954}. He, nevertheless, expected high activity in 2014.
Group G was identified with the $\zeta$ Draconids of \citet{Lindblad1995}.

The outburst predicted for 2014 really occurred and was observed visually and by radar, video, and photographic techniques 
\citep{RendtelMolau, Moorhead}. \citet{RendtelMolau} found that the enhanced activity was produced by radiants corresponding
to $\kappa$ Cygnids and $\alpha$ Lyrids defined by \citet{Lindblad1995}, not by his $\zeta$ Draconids. This result is consistent 
with the analysis of the 2007 outburst by \citet{Koseki2014}. \citet{Moorhead} presented 21 good quality photographic orbits 
from the Czech part of the European Fireball Network and analyzed them together with video data from two 
north American networks and radar data from the Canadian Meteor Orbit Radar (CMOR). The orbits were characterized by
wide range of semimajor axes with concentrations near the 5:3, 2:1, 9:4, and 3:1 resonances with Jupiter. The 5:3
resonance with 7.116 year period, corresponding to the apparent shower activity period, was discussed in more detail.
Only relatively small fraction of the dataset fell into this resonance so it seemed that it played only small role in the outburst. 
The radar activity in 2014 was found to be five times higher than in 2007 and nearly ten times higher 
than in all other years between 2002 and 2013. Video observations in 2015 showed the activity at background level again
\citep{Molau2015}.

\citet{CAMS} analyzed many meteor showers from the video observations by the Cameras for Allsky Meteor Surveillance (CAMS) project.
The data for $\kappa$ Cygnids cover years 2011 and 2012 when no outburst occurred. An extended band of
radiant concentrations, bent in the RA -- Declination plot, was observed. The activity moved from the eastern to the western part during August. 
The authors divided the meteors into four showers, $\kappa$ Cygnids (12/KCG), August Draconids (197/AUD), 
August $\mu$ Draconids (470/AMD), and $\iota$ Draconids (703/IOD), 
but noted that the division may not be unique. August Draconids were identified from radar data already
by \citet{Sekanina1976}. August $\mu$ Draconids were first defined on the basis of six meteors found by cluster analysis in the SonotaCo and CAMS
video databases by \citet{Rudawska}. They were also found in the European video database (EDMOND) by \citet{Kornos}.

\citet{RendtelArlt} reanalyzed the visual observation reports from the years 1975--2015 and found clear activity enhancement only in 1985
and 2014. Hints of higher activity were present in 1978, 2008, and 2011, while the outbursts observed  
by other techniques in 1993 and 2007 were not apparent in visual data.

\citet{Shiba2017} analyzed video data of the SonotaCo network from 2007--2016. Based on radiant position, he distinguished only two
showers: $\kappa$ Cygnids (which included also $\alpha$ Lyrids) and August Draconids (which included also
August $\mu$ Draconids and $\iota$ Draconids). Later,
\citet{Shiba2022} included also $\nu$ Draconids, discussed by \citet{CAMS2}, as a part of August Draconids.
The $\kappa$ Cygnids showed eight times enhanced activity over normal years in both 2007 and 2014.
The activity in 2013 was found to be enhanced two times and occurred at lower solar longitudes. The orbital periods of $\kappa$ Cygnids
were generally longer than corresponding to the 5:3 resonance but suffered from observational errors.

\citet{Koseki2020}, as a preparation for the expected return in 2021, studied the radiants in the $\kappa$ Cygnids region in three video
databases. Besides $\kappa$ Cygnids themselves, he identified two other showers. The first one was named August $\xi$ Draconids by him 
(the name was not recognized by the IAU MDC)
and was identical to Group F (also designed KCG3) in his previous work \citep{Koseki2014}. The second one was called $\zeta$ Draconids
following the earlier nomenclature of \citet{Lindblad1995} (distinct from the 73/ZDR)
and was identical to August Draconids (197/AUD) of \citet{CAMS}.

The high activity of kappa Cygnids in 2021 was reported already on August 9 in a telegram by \citet{CBET2021}. The outburst was observed visually
and by video \citep{Miskotte2022}.

\section{Our goal and methods}
\label{methods}

It is obvious that many open questions about  $\kappa$ Cygnids remain. There are strong suggestions that a meteoroid swarm
in 5:3 resonance in Jupiter is responsible for periodic enhancements of activity. That hypothesis was, however, not directly proven by measurements
of orbital periods of the meteoroids. Little is known about the annual component of  $\kappa$ Cygnids and
its relation to other showers active in August with similar radiants. There is even no consensus about the number of these similar showers.
In his latest book, \citet{JennBook2}, following \citet{Shiba2017, Shiba2022}, considers only August Draconids. On the other hand, 
the IAU MDC\footnote{https://www.iaumeteordatacenter.org/ accessed October 2, 2024} keeps more showers.
August Draconids (197 AUD) are listed as an established shower,  August $\mu$ Draconids (470 AMD) and $\nu$ Draconids (220 NDR)
are marked as nominated to be established, and $\iota$ Draconids (703 IOD) are on the working list. 

Our goal is to study $\kappa$ Cygnids and other nearby showers independently using new data from the European Fireball Network (EN).
EN data were already used for the study of $\kappa$ Cygnids in 2014 \citep{Moorhead}. At that time, the analog Autonomous Fireball
Observatory (AFO) was used as the main instrument but the new Digital Autonomous Fireball Observatory (DAFO) was already installed at six
stations. As of 2024, DAFOs, which are more sensitive and more efficient, are used as the main instrument at 20 stations. DAFOs take two high
resolution digital images of the whole sky per minute and are equipped with LCD shutters to measure meteor velocities. They contain also
radiometers which provide high resolution light curves of brighter meteors with up to 5000 samples per second. More details about the network and
the used data procedures can be found in \citet{Taurids} and \citet{catalog}.

 \begin{figure}
    \centering
    \includegraphics[width=\linewidth]{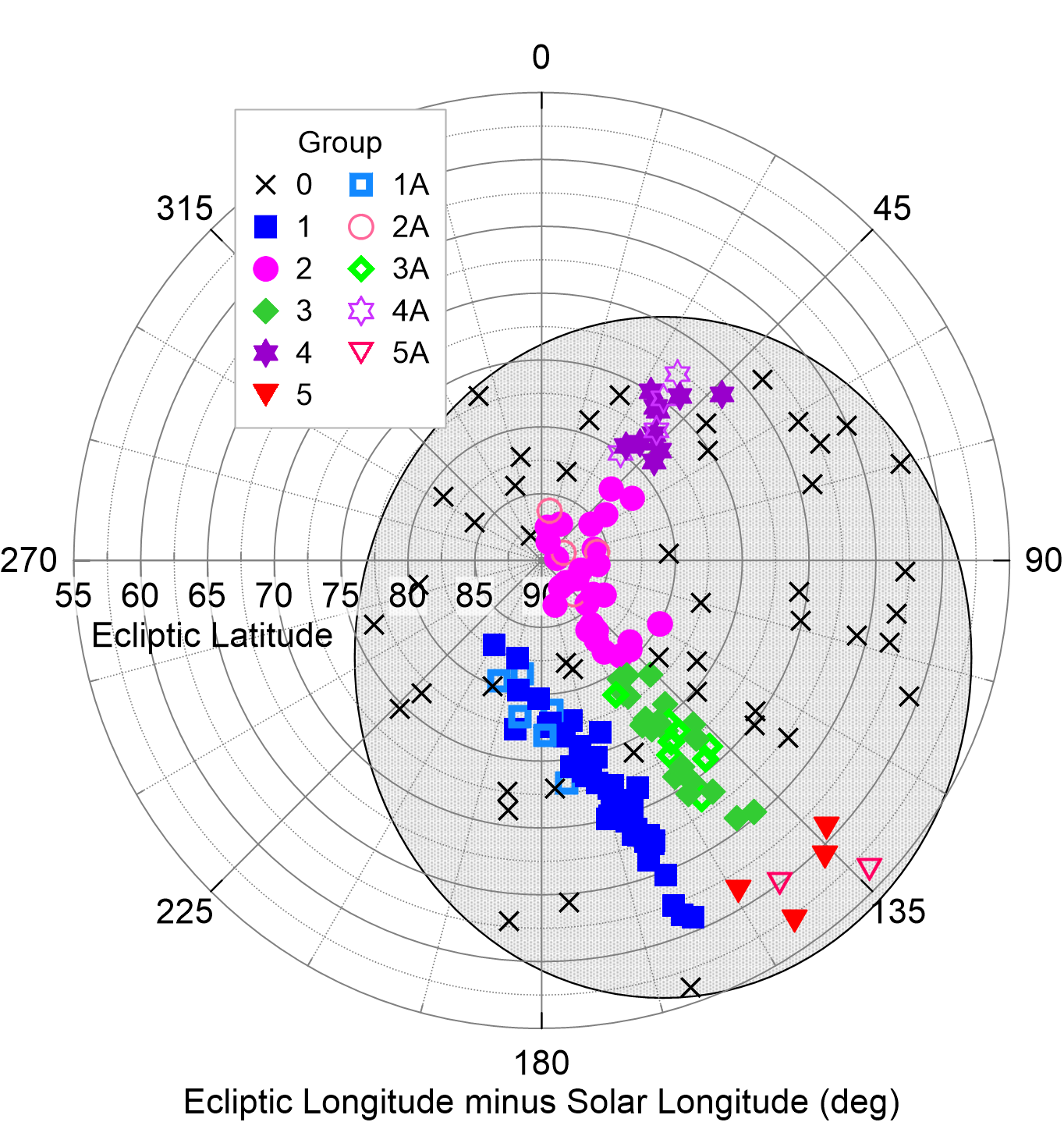}
    \caption{Positions of geocentric radiants of the analyzed meteors in Sun-centered ecliptical coordinates in polar projection.
    The shaded area is the investigated region of interest. The radiants of meteors which were assigned to one of five groups
    are plotted by color symbols. Other radiants are plotted as crosses.}
    \label{radiants-polar}
\end{figure}

 \begin{figure}
    \centering
    \includegraphics[width=\linewidth]{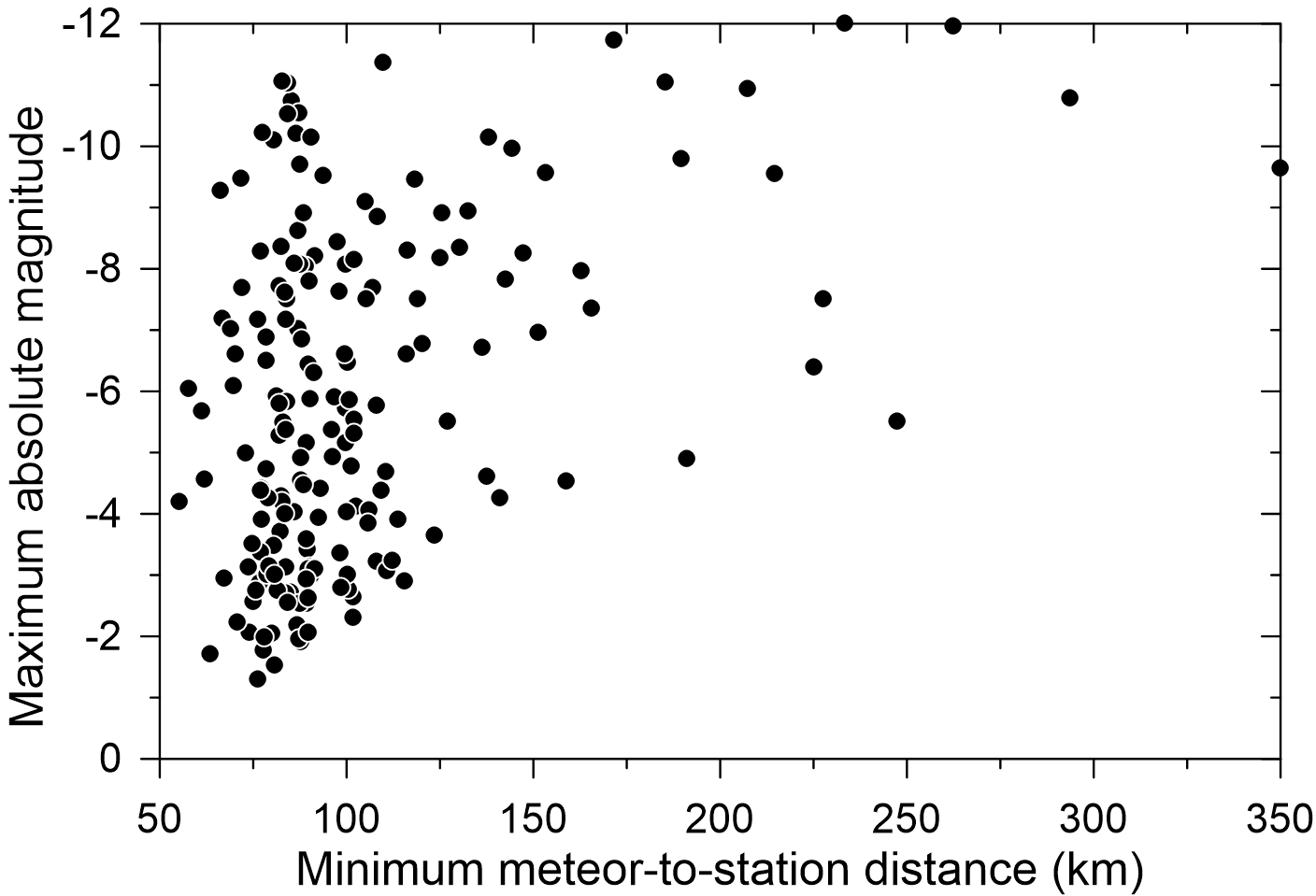}
    \caption{Meteor maximum absolute magnitude plotted against the distance of the meteor from the nearest observing station.}
    \label{magnitudes}
\end{figure}

The data were analyzed without previous assumptions about meteor showers, just selecting the period of activity and broad radiant area. All
meteors observed in the month of August in the years 2016--2024 and having radiants in the region of interest were considered.
The region of interest included western Cygnus, southern Draco, Lyra, and eastern Hercules. It is plotted in ecliptic coordinates in
Fig.~\ref{radiants-polar}. 
Meteors too short for DAFOs to be included in our catalog \citep{catalog} were also considered.
Additional measurements were done on images from SDAFOs, the spectral versions of DAFOs,
and on videos from the supplementary video cameras (IP cameras). SDAFOs are more sensitive than DAFOs since they use lenses
with longer focal lengths and do not use LCD shutters. They can be therefore used to improve the trajectory and radiant solution.
Even more useful are the IP cameras. Because of their higher sensitivity (in comparison with DAFO), they capture meteors earlier and 
can be used to improve the solution of initial velocity. IP cameras started to be deployed in 2016. Nowadays, there are batteries of them covering
together all sky at two stations and each other station has at least one IP camera with typical field of view of $100\degr \times\ 55\degr$.

The limiting sensitivity of the network depends on the position of the meteor. Figure~\ref{magnitudes} shows that 
meteors with maximum magnitude $-2$ or even
fainter could be analyzed if they were close to one of the stations (i.e.\ visible near zenith). Bright meteors (fireballs)
could be observed also if they were further away. The brightest recorded fireballs reached the magnitude $-12$, typically in a flare. 

The total number of analyzed meteors is 179. In the following section, we will try to identify groups of meteors belonging to the same meteor shower
on the basis of radiant position and elements of the heliocentric orbit.

\section{Defining meteor groups}
\label{groups}

The radiant plot in Fig.~\ref{radiants-polar} shows that radiants are not distributed randomly within the region of interest.
We tentatively defined five meteor groups. Although the radiant position was the primary criterion, some boundary cases
were decided according to additional criteria described below. 

Considering the radiants, Group~1 (blue in Fig~\ref{radiants-polar}) is relatively well defined although the radiant area is very
elongated, spanning more than 20\degr\ in ecliptic latitude. Groups 2, 3, and 4 also correspond to obvious radiant concentrations
but are not well separated each from other. Group~5 is a rather spurious radiant concentration and contains only six members. Meteors
marked as Group~0 do not belong to any of the five groups and are probably sporadic.

 \begin{figure}
    \centering
    \includegraphics[width=\linewidth]{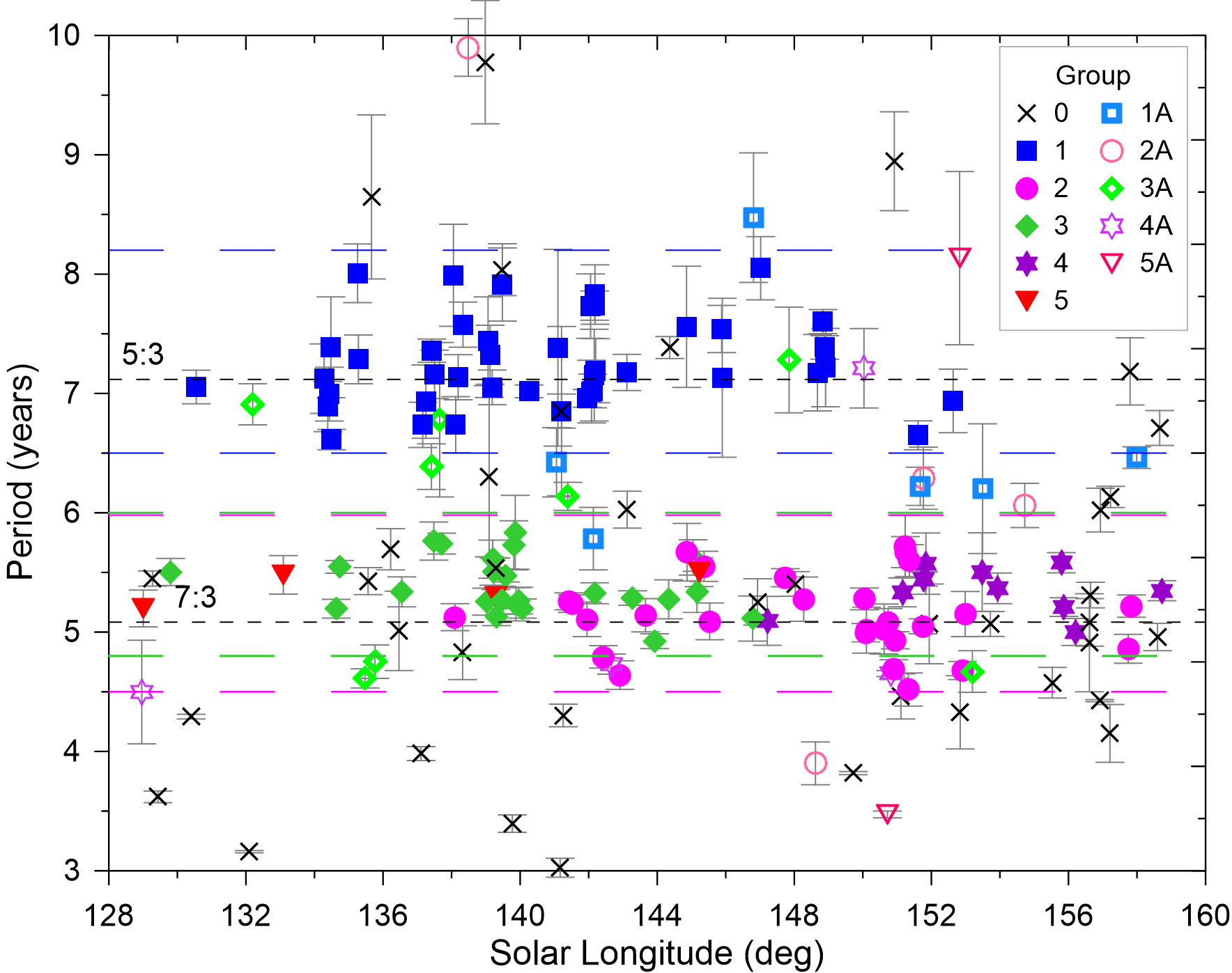}
    \caption{Orbital periods as a function of solar longitude. Formal one-sigma errors of periods are indicated.
     Only periods within 3 -- 10 years are shown. Seven meteors had periods outside
    this range; the full range was 1.3 -- 15 years. Solar longitudes correspond to the month of August. Long-dash color lines show the
    limits of periods for the cores of individual groups. Short-dash black lines mark the 5:3 and 7:3 resonances with Jupiter.}
    \label{periods}
\end{figure}

 \begin{figure}
    \centering
    \includegraphics[width=0.9\linewidth]{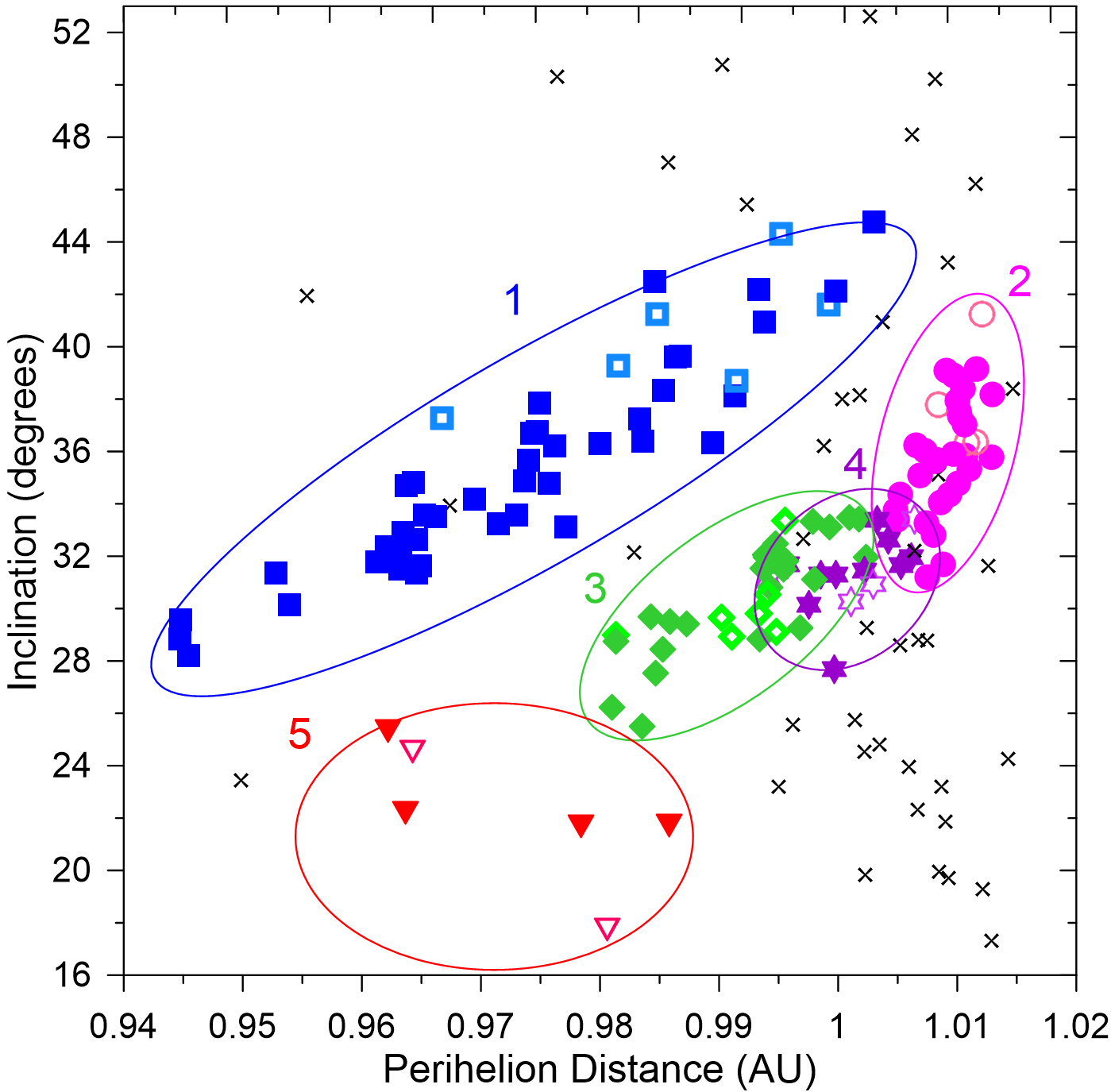}
    \caption{Relation between perihelion distance and inclination. Groups are coded as in Fig.~\ref{radiants-polar} 
    and marked by color ellipses.}
    \label{q-i}
\end{figure}

 \begin{figure}
    \centering
    \includegraphics[width=0.9\linewidth]{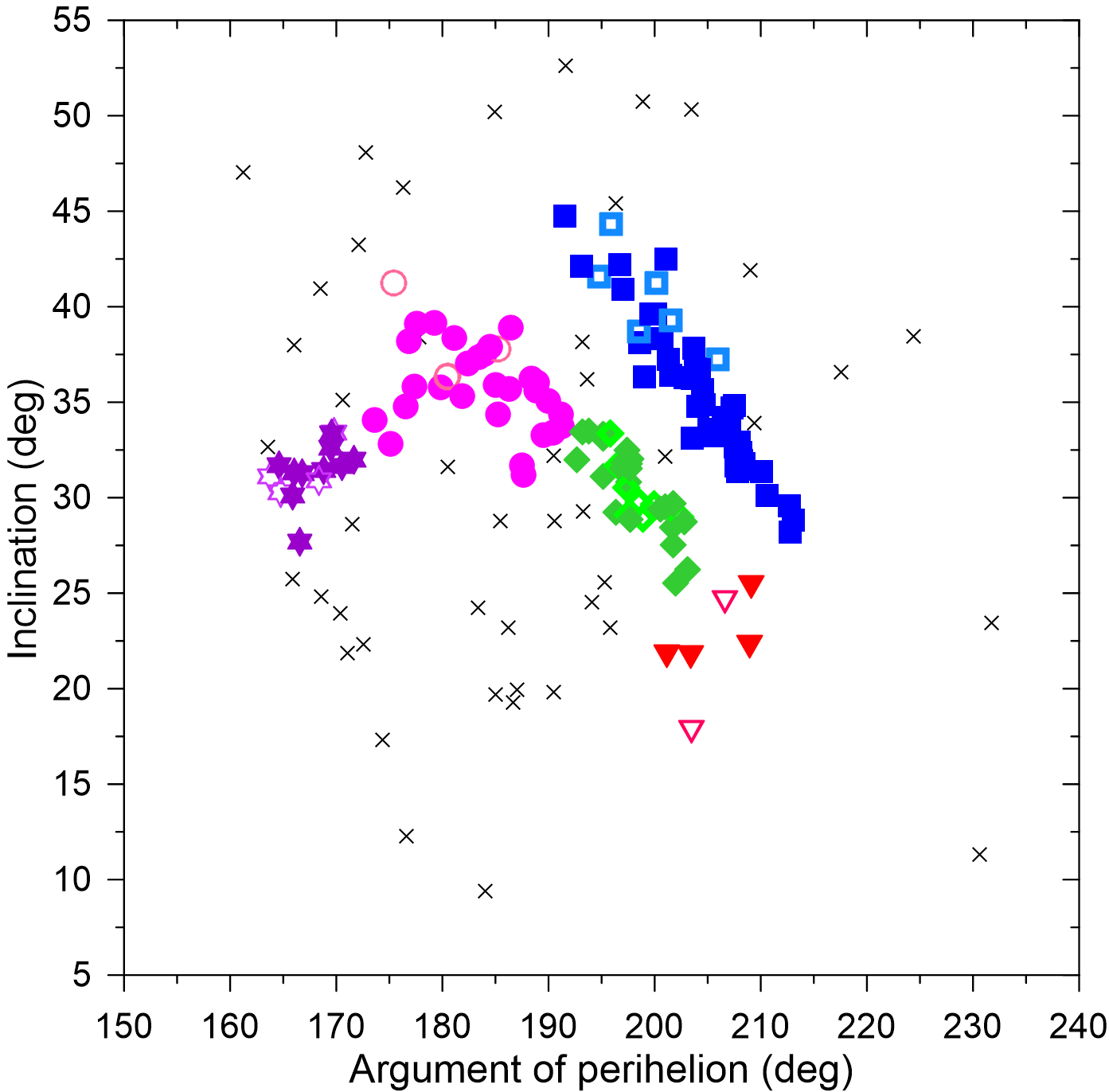}
    \caption{Relation between argument of perihelion and inclination. Groups are coded as in Fig.~\ref{radiants-polar}.}
    \label{argup-i}
\end{figure}

 \begin{figure*}
    \centering
    \includegraphics[width=0.42\linewidth]{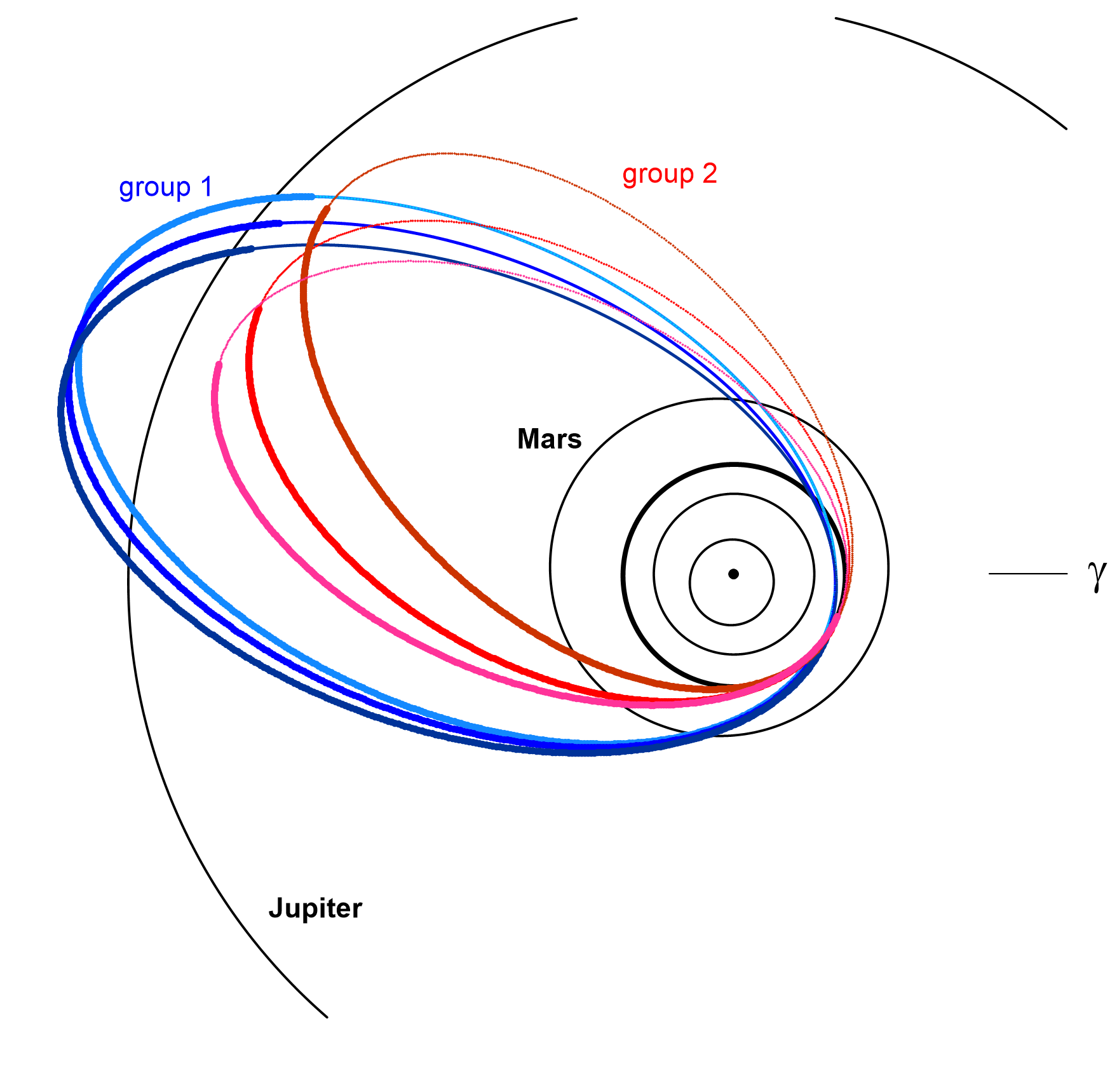}
    \hspace*{0.08\linewidth}
    \includegraphics[width=0.42\linewidth]{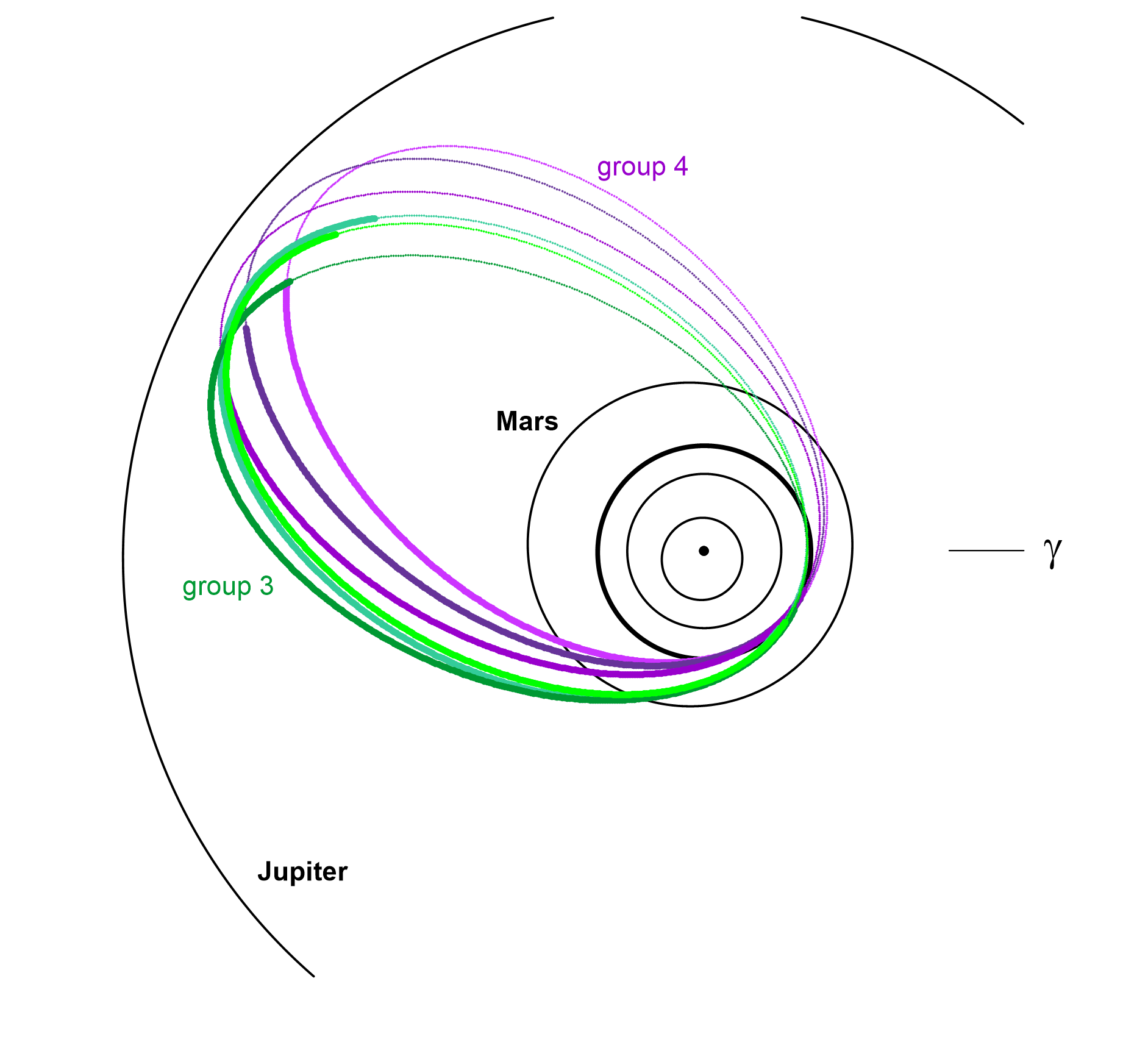}
    \caption{Projections of typical meteoroid orbits from Groups 1 and 2 (left) and 3 and 4 (right) into the ecliptic plane.
    The parts of orbits above ecliptic are shown by thick line. Orbits of planets Mercury to Jupiter are drawn in black, the Earth's
    orbit is thicker. Vernal equinox is to the right. All planets and meteoroids move counterclockwise. Note the different positions 
    of nodes relative to aphelion in Groups 2--4.}
    \label{orbits}
\end{figure*}

The next criteria for group assignment are the day of appearance, expressed as solar longitude, and the semimajor axis or,
alternatively, orbital period.  Figure~\ref{periods} shows the period as a function of solar longitude. No concentration in solar longitude
can be seen. Nevertheless, Group~3 was more active in the first half of August and Group~4 was active only in the second half of August.
Since we limited our analysis to meteors observed in August, we have possibly missed part of Group~4, which may be
active at the beginning of September.

Group~1 is clearly separated from other groups in orbital period. We defined the core of each group by a range of orbital periods.	
For Group~1, is was 6.5--8.2 yr, for Group~2, 4.5--6 yr, and for Groups 3--5, 4.8--6 yr. Meteors falling outside these ranges are marked as
Groups 1A to 5A and are plotted by open symbols in all graphs. They may be either normal members of the groups with inaccurately determined
period, or scattered members with modified period, or random interlopers not belonging to the group. We cannot distinguish
between these possibilities.

An interesting pattern occurs when plotting perihelion distance and inclination (Fig.~\ref{q-i}). There is a clear correlation
between these two quantities for Group~1. As the perihelion distance increases from 0.94 to 1 AU, the inclination
also increases from 28\degr\ to 44\degr. This plot can be used to identify possible members of Group~1. Groups 2--5 together cover
similar range of perihelion distances but the inclination for a given perihelion distance is always lower. Group~4 overlaps with 
Groups~2 and 3 in this plot.

An alternative plot is between argument of perihelion and inclination (Fig.~\ref{argup-i}). All groups can be separated. 
This plot reflects to some extent the radiant distribution in  Fig.~\ref{radiants-polar}.

\section{Understanding the geometry of the orbits}
\label{understanding}

The orbits of all groups have perihelia close to the Earth orbit (0.94 -- 1.015 AU) and moderate inclinations 20\degr\ -- 45\degr.
The meteoroids encounter the Earth in the descending node when crossing the ecliptic from top to bottom. The projections
of representative orbits from Groups 1--4 into the ecliptic plane are shown in Fig.~\ref{orbits}. Group~5 is similar to Group~3
in this projection.

The aphelia of Group~1 lie well behind the orbit of Jupiter. The ascending node is close to the Jupiter's orbit. The crossing
with the Earth's orbit occurs before reaching perihelia. The perihelion points lie below the ecliptic.

All other groups have aphelia inside of the Jupiter's orbit but relatively close to it. Group~2 has the ascending nodes at aphelia
and descending nodes at perihelia. The perihelia lie very close to the Earth's orbit. Since the meteoroid radial (toward Sun) speed
is zero during the Earth encounter, and the longitudinal (along Earth's orbit) speed is close to the Earth's orbital speed, the geocentric
radiants are very close to the north pole of ecliptic.

Groups~3 and 4 are nearly symmetrical. The perihelia lie just inside of the Earth's orbit in both cases and the inclinations are nearly 
the same.  Group~3 has the ascending nodes before aphelia and descending nodes before perihelia.
The perihelion points lie below the ecliptic. Group~4 has the ascending nodes after aphelia and descending nodes after perihelia.
The perihelion points lie above the ecliptic.

 \begin{figure}
    \centering
    \includegraphics[width=0.8\linewidth]{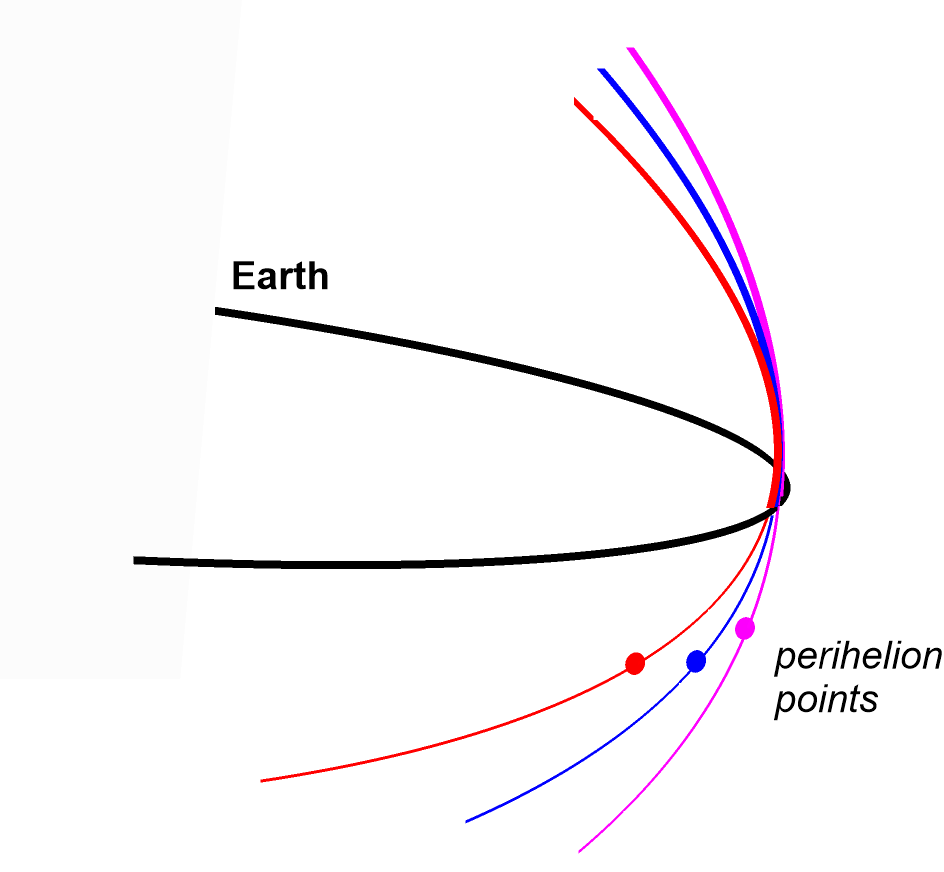}
    \caption{Three representative orbits of Group~1 meteoroids in a three-dimensional plot with Earth's orbit. The positions
    of perihelion points are depicted as circles. The parts of orbits below ecliptic are shown by thin line.}
    \label{sideview}
\end{figure}

To explain the correlation between perihelion distance, $q$, and inclination, $i$, in Group~1, three orbits with different $q$
are plotted in Fig.~\ref{sideview} in a side view. The perihelion points lie inside of the Earth's orbit below the ecliptic. 
To intersect Earth's orbit and thus be observable, the orbits with larger $q$ must have larger $i$
than those with lower perihelion distances. The observed $q$-$i$ correlation thus seems to be a selection effect from the condition of
Earth intersection. Orbits with other combinations of $q$ and $i$ probably exist in the stream but cannot be observed on Earth.

There is a smooth transition between Groups 2, 3, and 4. All groups have similar semimajor axes and are probably parts
of a single meteoroid stream. Group~2 has a slightly higher inclinations than the other two groups (Fig.~\ref{q-i}) 
but this may again be a geometric condition of Earth intersection. Group~5 is less compact and not well represented in our sample
but may also be part of the same stream. Its semimajor axis is similar to Group~3 and it is close to Group~3 in Figs.~\ref{q-i} and \ref{argup-i}.

 \begin{figure}
    \centering
    \includegraphics[width=1.0\linewidth]{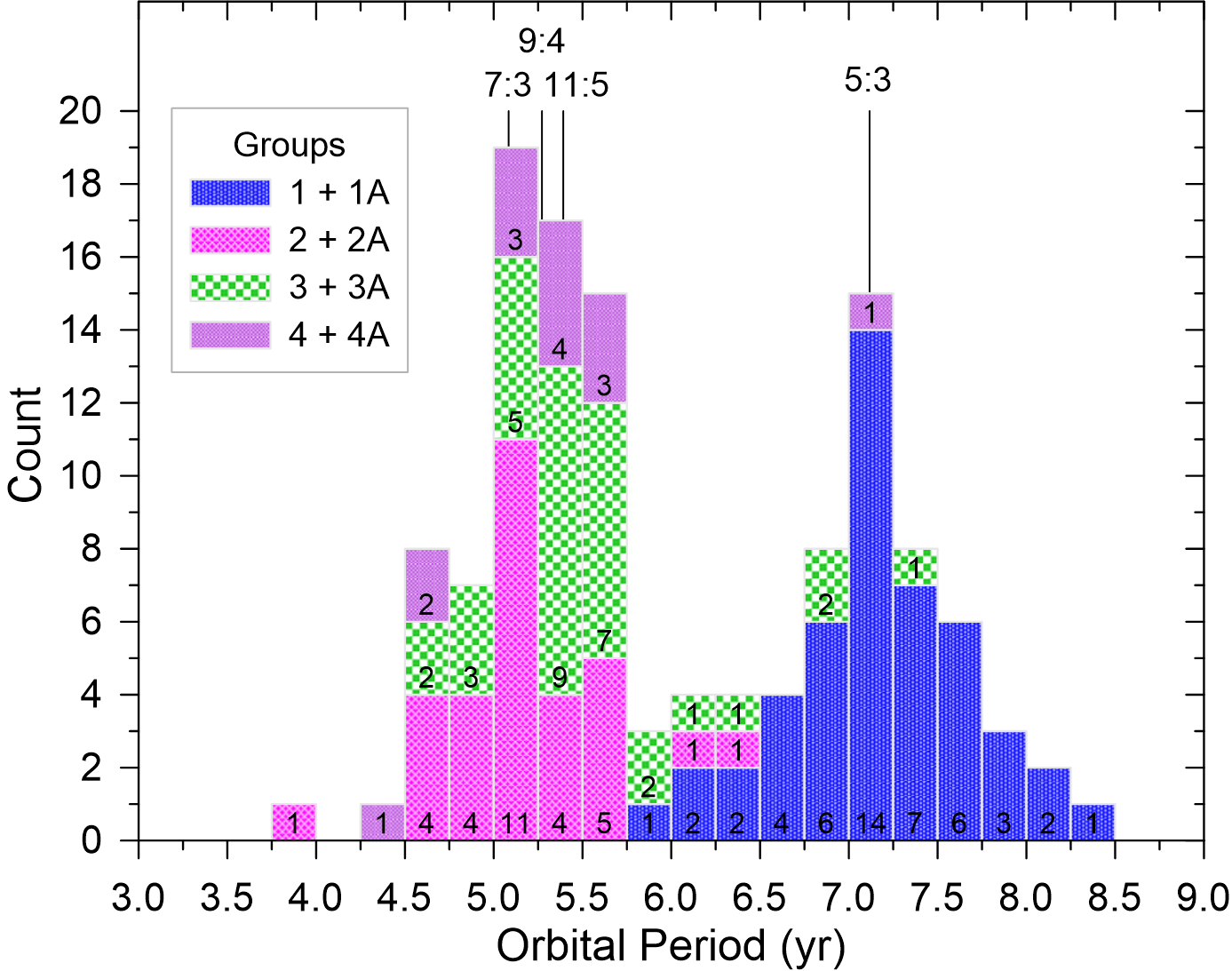}
    \caption{Histogram of orbital periods of meteoroids of Groups 1 to 4, including the outliers. Positions of selected mean motion resonances
    with Jupiter are also indicated.}
    \label{hist-period}
\end{figure}

\begin{table}
\caption{Average orbital periods of individual groups without outliers.}
\label{table-orbit}
\begin{tabular}{llll}
\hline
Group &Period & Number  & Possible  \\
&years &of orbits & resonances\\
\hline
1 & 7.26  $\pm$  0.38 & 42 & 5:3 \\
2 & 5.14  $\pm$  0.32 & 28 & 7:3, 9:4 \\
3 & 5.40  $\pm$  0.24 & 24 & 9:4, 11:5 \\
4 & 5.35  $\pm$  0.20 & 10 & 9:4, 11:5 \\
\hline
\end{tabular}
\end{table}

\section{Orbital periods and resonances}

Determining precise semimajor axes and orbital periods of cometary meteoroids from meteor observation is a challenge. 
Small errors in the measurement of the entry speed result in large errors of semimajor axes. Since fragile cometary meteors
usually have short atmospheric trajectories, precise speed determination is difficult. Previous studies failed to confirm the
suspected orbital period of $\kappa$ Cygnids of $\sim$7 years, which was based on the periodicity of the activity. Our data are
more precise but still there are appreciable uncertainties of orbital periods of individual meteoroids.

In Fig.~\ref{hist-period}, a histogram of orbital periods of meteoroids of Groups 1 to 4 is given. No period cutoff was applied,
so all meteoroids falling into the group on the basis of the radiant position are included in the plot. Group~5 was omitted
because of small number of meteoroids.

As previously, we can see clear separation of Group~1 from other groups. There is a nice distribution of periods of Group~1 with the
peak corresponding to the 5:3 resonance with Jupiter (7.12 yr). The average periods for all groups are given in Table~\ref{table-orbit}.
Only the cores of the groups were taken into account here. When including outliers, the mean orbits are almost the same, 
but the errors are much larger (7.18  $\pm$  0.52 yr  for Group~1; for other groups the errors are between 0.6 and 1 year).

The periods of Groups 2--4 (and also Group~5) are the same within the uncertainties. They may fall into one or more higher order resonances
with Jupiter, namely 7:3 (5.08 yr), 9:4 (5.27 yr), or 11:5 (5.39 yr) but it is not certain if the orbits are indeed resonant. 
There is no evidence for a peak at the 2:1 resonance (5.93 yr) and also the 5:2 resonance (4.75 yr) is probably absent.

 \begin{figure}
    \centering
    \includegraphics[width=1.0\linewidth]{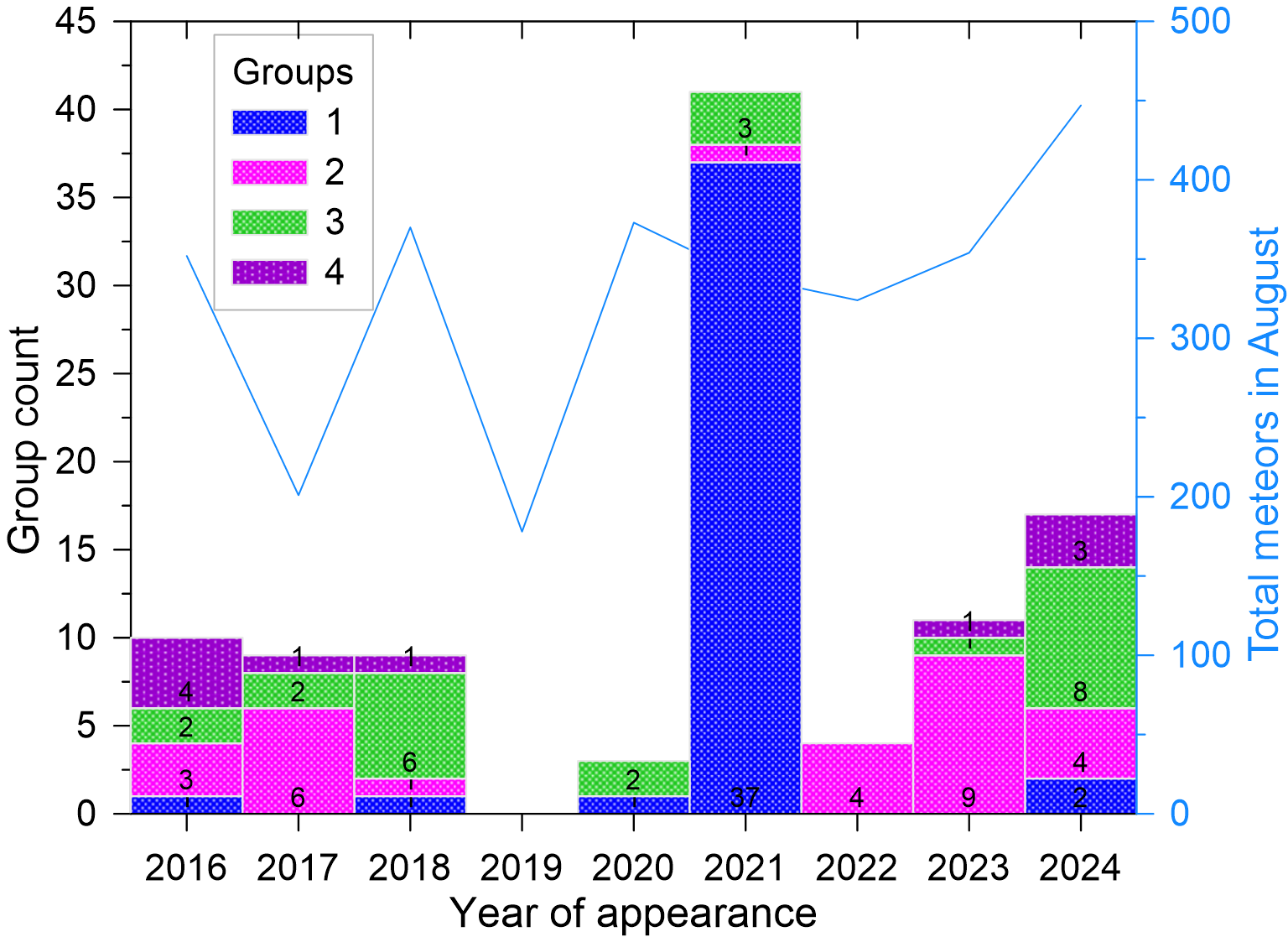}
    \caption{Histogram of year of appearances of meteors belonging to the cores of Groups 1 to 4. The four
core members of Group~5 are not shown. They appeared in four different years (2020-22, 2024). The solid blue line denotes
    the total number of meteors observed in August each year. The corresponding scale is on the right.}
    \label{hist-years}
\end{figure}

 \begin{figure}
    \centering
    \includegraphics[width=0.9\linewidth]{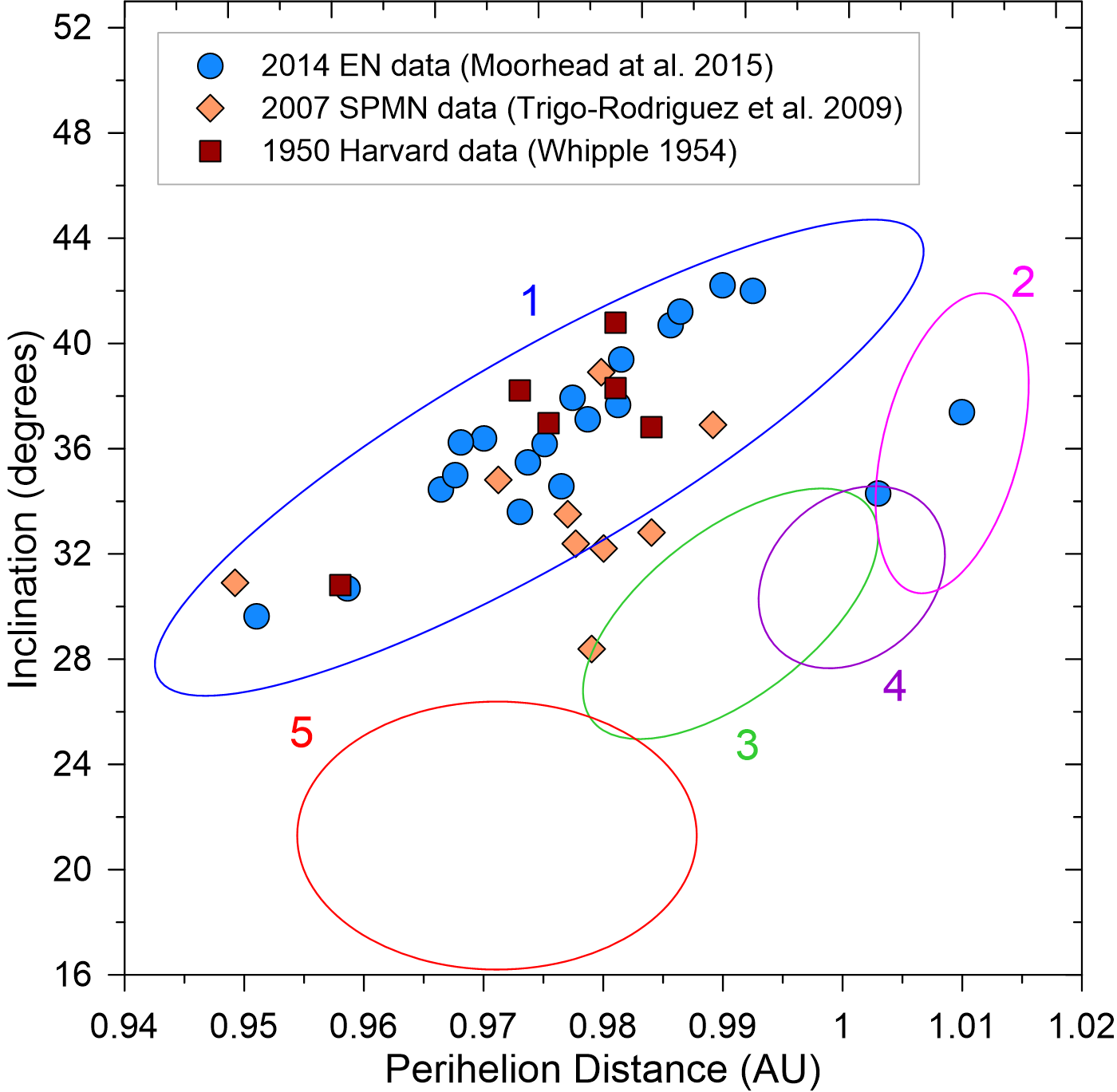}
    \caption{Relation between perihelion distance and inclination for the 2014 $\kappa$ Cygnid data from the Europaen Fireball Network,
    2007 data from the Spanish Meteor Netwotk, and 1950 data from Harvard cameras. Groups are marked as in Fig.~\ref{q-i}. 
    One 2014 fireball with $q=0.916$ AU is not plotted.}
    \label{2014data}
\end{figure}

 \begin{figure*}
    \centering
    \includegraphics[width=0.8\linewidth]{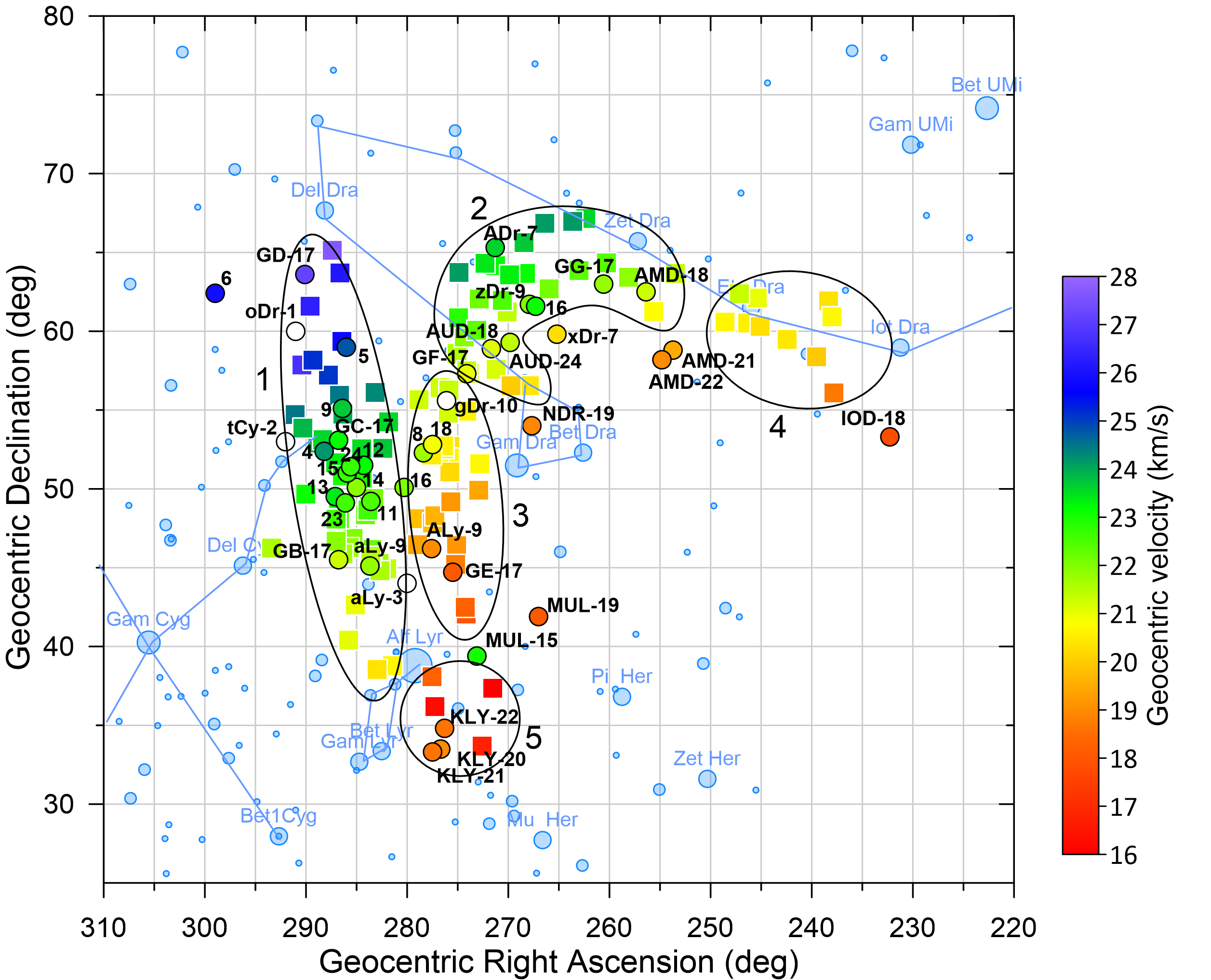}
    \caption{Geocentric radiants of meteors belonging to the cores of Groups~1--5 (color squares) plotted on the background sky map.
    Geocentric velocity is color coded. The approximate limits of the groups are plotted as black ellipses or other shapes. Plotted are
    are also radiants of various showers from the literature (color circles) with velocities (if known) coded with the same colors. 
    The showers are labeled according to the names used by the authors and the paper numbers as given 
    in Tables~\ref{KCGdata}--\ref{G4data} (number 19, not listed in the Tables, belongs to \citet{CAMS2}).
    Only paper numbers are given for $\kappa$ Cygnids. The other shower names are as follows: 
    tCy ($\theta$ Cygnids) oDr ($o$ Draconids), zDr ($\zeta$ Draconids), gDr ($\gamma$ Draconids),
    xDr ($\xi$ Draconids), ADr or AUD (August Draconids), AMD (August $\mu$ Draconids), IOD ($\iota$ Draconids), NDR ($\nu$ Draconids),
    aLy ($\alpha$ Lyrids), ALy (August Lyrids), MUL ($\mu$ Lyrids), KLY ($\kappa$ Lyrids), GB to GG (Groups B to G of Cygnid-Draconid Complex).}
    \label{RA-Dec}
\end{figure*}

\section{Periodic activity}
\label{periodicity}

The numbers of meteors of each group observed in individual years are shown in Fig.~\ref{hist-years}.
To be sure that only true members of the showers are considered, the outliers were excluded.  It is immediately
obvious that the activity of Group~1 occurred almost exclusively in 2021. From 42 meteors classified as the core of this
group, 37 appeared in 2021. Two were observed in 2024 and three in other years. All these three lie rather close to the
boundary of Group~1 in Fig.~\ref{q-i}, so they were not typical members of the group. From six outliers in this group
with orbital periods outside the selected range
(not shown in Fig.~\ref{hist-years}), five appeared in 2021, suggesting that they were related to the group.

The prevalence of the year 2021 in Group~1 is consistent with the 7~year periodicity of $\kappa$ Cygnids. The previous years
of high activity were 2014 and 2007 (see Sect.~\ref{history}). We plotted the $q$-$i$ graph
for the 2014 EN data \citep{Moorhead}, 2007 Spanish data \citep{Trigo2009}, and 1950 Harvard data \citep{Whipple1954}
in Fig.~\ref{2014data}. From 21 fireballs
observed in 2014, 18 nicely fit into the Group~1 region. Group~1 was therefore responsible for the enhanced activity in 2014. 
The fit is not so nice for the 2007 data, probably due to a lower quality of these data. Still, most data points lie
within or close to the Group~1 region and we can conclude that this group was also responsible for the 2007 activity.
We do not compare semimajor axes since they are less reliable in older works.
The five 1950 orbits fall into Group~1 well. The year 1950 does not align with 7-year periodicity but the period is in fact slightly
larger than 7 years, see Sect.~\ref{periodicity_discus}.

None of other groups showed such concentration as Group~1 (Fig.~\ref{hist-years}). There is a hint of a fluctuation with a minimum in 2019--2020
but the statistics is poor. Moreover, the overall number of detected meteors in August 2019 was lower than in other years due to worse
observing conditions. Year-to-year variations in activity of Groups 2--4 can be therefore neither confirmed nor ruled out at this stage.

\section{Identification with previously reported showers}
\label{identification}

In this section we identify our Groups~1--5 with meteor showers reported previously in the literature under various names. We use primarily
the radiant positions and velocities. Because of very large sizes of radiant areas and low precession rate near the ecliptic pole, 
we ignore the differences between different equinoxes and take the mean shower radiants directly as published in original papers.

The radiant positions and velocities are compared graphically in Fig.~\ref{RA-Dec}. We can see that the observed meteor radiants and the
reported showers largely overlap. Also the velocities are consistent in most cases. The radiant derived for $\kappa$~Cygnids by most authors
overlaps with the center of Group~1. Only the radar determined radiant of \citet{Sekanina1973} is off.
The radiant area of Group~1 is very extended in declination 
and encompasses also $o$ Draconids of \citet{Denning1879a} to the north
and $\alpha$ Lyrids of \citet{Lindblad1995} to the south. Groups B to D of \citet{Koseki2014} are also parts of our Group~1.
There is a clear correlation between declination and velocity. It is investigated in the next section.

The transition between Groups 2, 3, and 4 is smooth not only in terms of radiant position but also in terms of velocity. Similarly to Group~1,
radiant positions and velocities are correlated. In contrast, while the radiants of Groups~1 and 3 are also close, there is a jump in velocity between them. 
This is another evidence that Groups~2--4 are branches of the same stream and Group~1 is a separate stream.

The radiant of $\kappa$ Cygnids reported by \citet{CAMS} falls within Group~3. It is not surprising because the authors used data from
2011--2012 when Group~1 was probably not active (or only marginally). Also the $\kappa$ Cygnid radiant of \citet{Porubcan} falls within Group~3.
They probably included members of both Groups 1 and 3 in their computer search. Group~3 further includes August Lyrids
of \citet{Lindblad1995}, Koseki's Group~E, and $\gamma$ Draconids of \citet{Jones2006}. 
Koseki's Group~F lies at the boundary of Groups~2 and 3. Group~2 contains
August Draconids (197/AUD), $\zeta$ Draconids of \citet{Lindblad1995} and Koseki's Group~G. Also the late activity of $\kappa$~Cygnids  observed in 2012 by
\citet{Molau2012} was in fact Group~2.  Finally, the August $\mu$ Draconids (470/AMD) reported by \citet{CAMS} fall within our
definition of Group~2. The radiants and velocities of the same shower found by \citet{Rudawska} and \citet{Kornos} lie
outside our groups but close to both Groups~2 and 4. There is no direct overlap of any radiant with Group~4. The $\iota$ Draconids (703/IOD)
lie farther in that direction. 

Group~5 corresponds well with $\kappa$ Lyrids (464/KLY). This shower, reported by \citet{Rudawska}, \citet{Kornos}, and \citet{CAMS3},
has the maximum of activity in July and was therefore not completely covered by our analysis. It will be therefore not discussed in more detail here.
Similarly, the $\mu$ Lyrids (413/MUL) and $\nu$ Draconids (220/NDR) \citep[= $\xi$ Draconids of][]{Sekanina1976} have similar types of orbits as our groups
but are mostly active in July and September, respectively, and could not be confirmed here.
However, it is clear that there are other showers of the proposed C-D Complex active in August besides the $\kappa$ Cygnids 
but they cannot be distinguished on the basis of radiant and velocity.
It was therefore legitimate that \citet{JennBook2}
kept only $\kappa$ Cygnids and August Draconids and removed all other showers mentioned here.

 \begin{figure}
    \centering
    \includegraphics[width=0.9\linewidth]{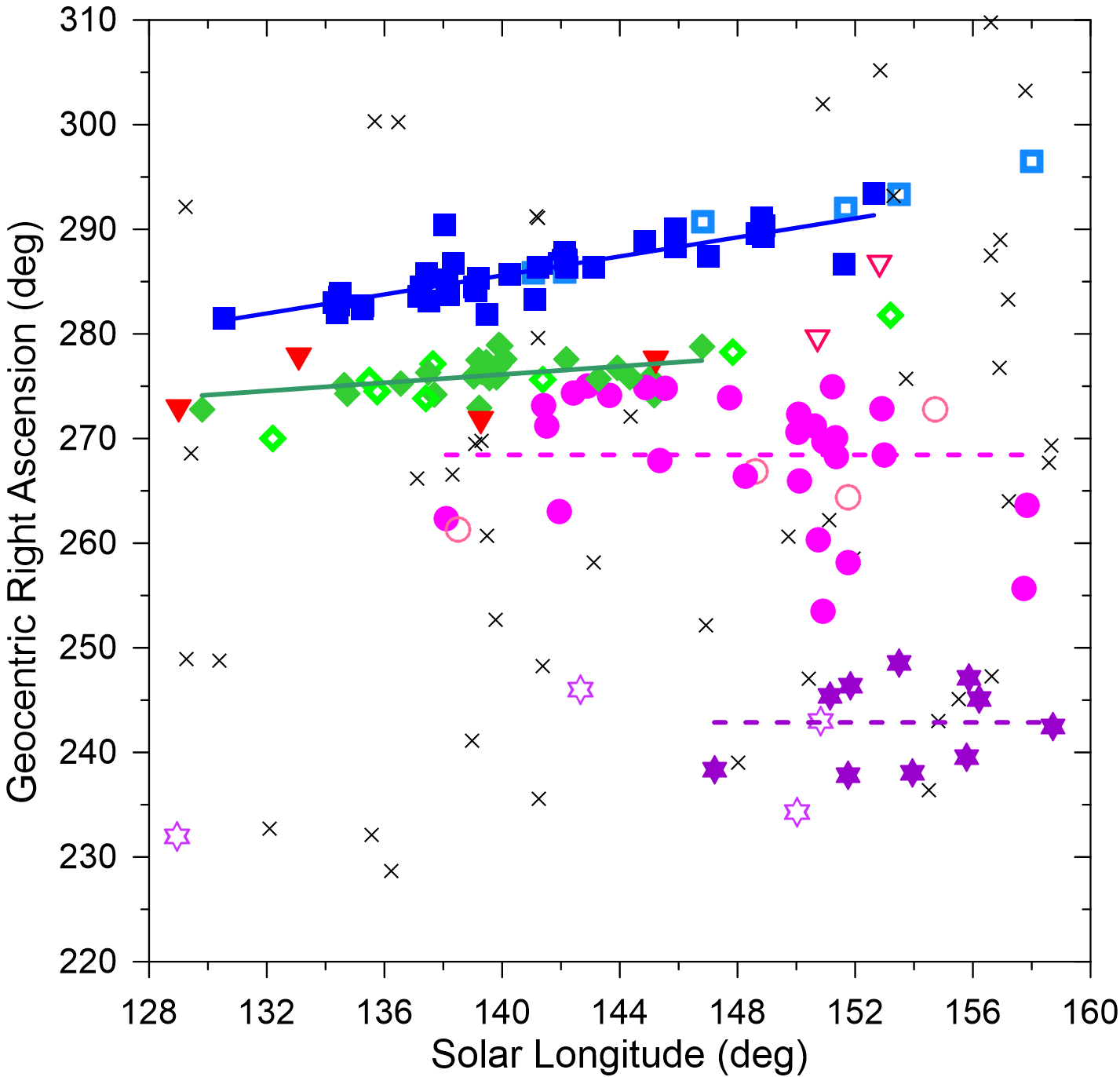}
    \caption{Geocentric right ascension of the radiant as a function of solar longitude. Groups are coded as in Fig.~\ref{radiants-polar}. 
    Linear fits are drawn for the core members of Groups~1 and 3 (solid lines). Mean values are plotted for Groups~2 and 4 (dashed lines).}
    \label{solong-RA}
\end{figure}

 \begin{figure}
    \centering
    \includegraphics[width=0.9\linewidth]{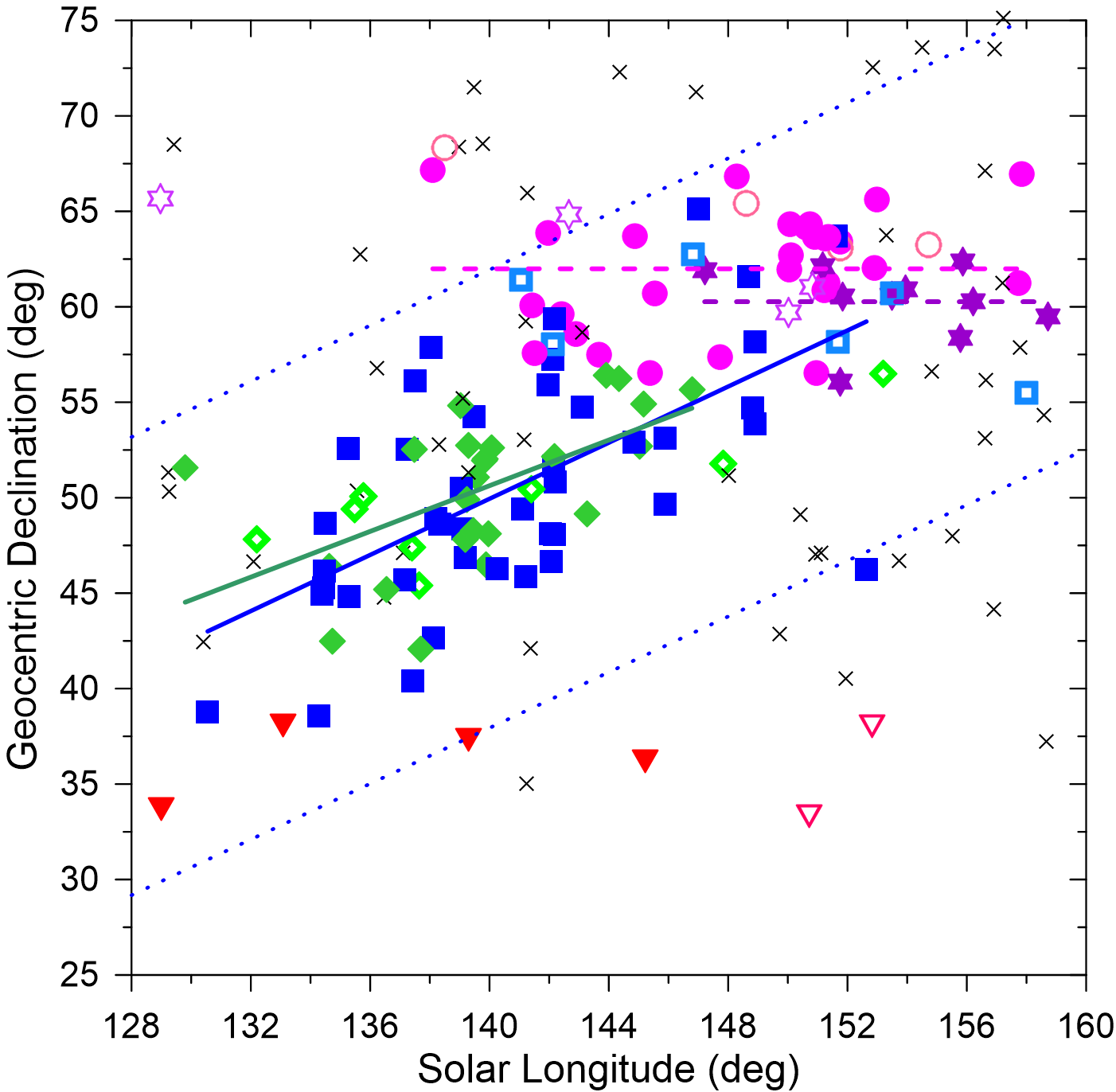}
    \caption{Geocentric declination of the radiant as a function of solar longitude. Groups are coded as in Fig.~\ref{radiants-polar}. 
    Linear fits are drawn for the core members of Groups~1 and 3 (solid lines). For Group~1, limits of $\pm12\degr$ in declination 
    from the fit are also drawn (dotted lines). Mean values are plotted for Groups~2 and 4 (dashed lines).}
    \label{solong-Dec}
\end{figure}

 \begin{figure}
    \centering
    \includegraphics[width=0.9\linewidth]{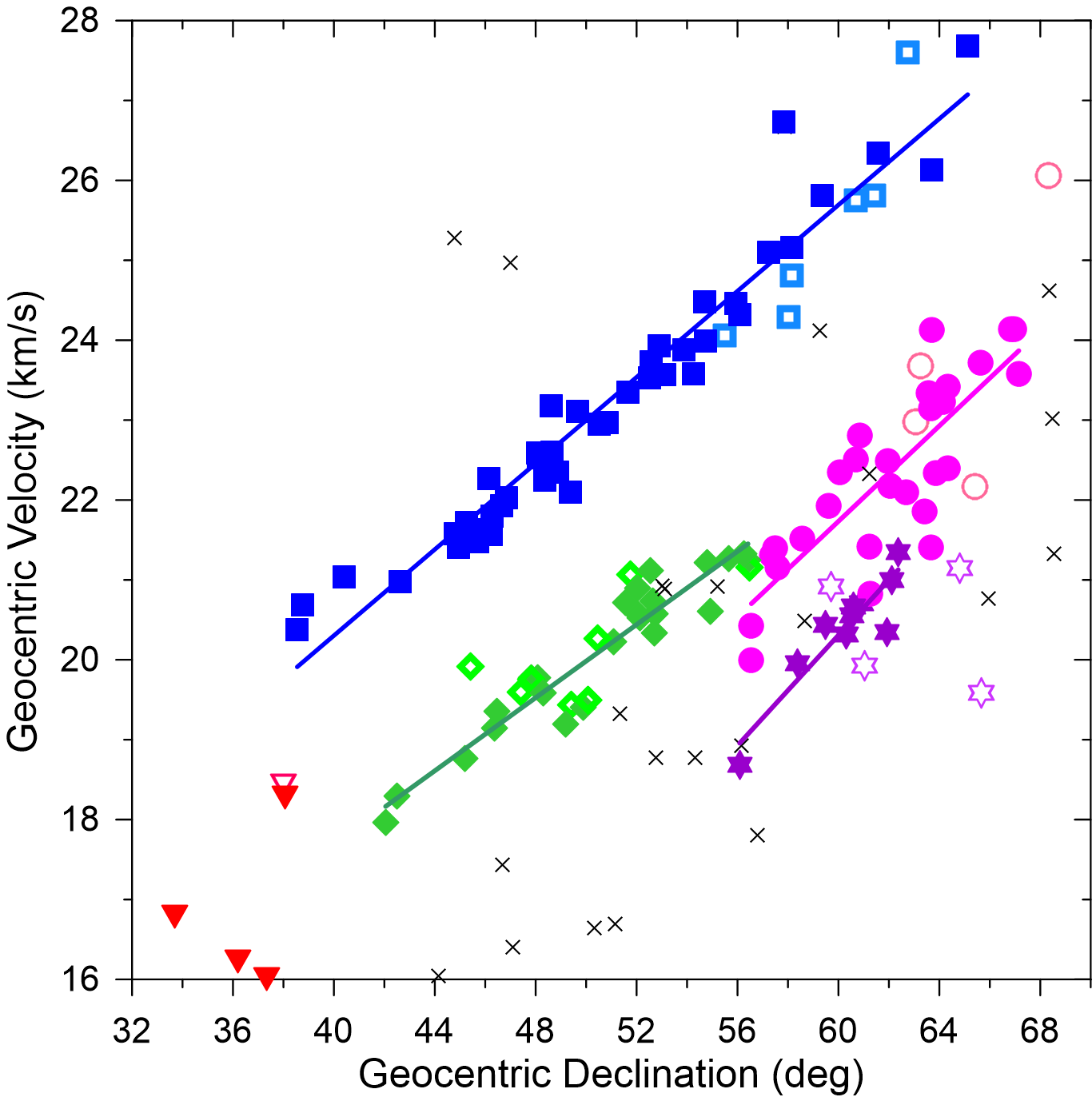}
    \caption{Correlation between geocentric declination and velocity. Groups are coded as in Fig.~\ref{radiants-polar}. 
    Linear fits are drawn for the core members of Groups~1--4 (solid lines).}
    \label{Dec-Vg}
\end{figure}

\begin{table*}
\caption{Five representative sets of $\kappa$ Cygnid (Group~1) radiants and orbits and identification of corresponding showers from the literature.}
\label{KCGdata}
\begin{tabular}{llllllllll} \hline
Designation & Reference & $\alpha_{\rm g}$ & $\delta_{\rm g}$ &  $v_{\rm g}$ &  $a$ & $q$ & $i$ & $\omega$ & $\Omega$ \\
\hline
$p=-0.8$& this work &286.5&  41.8&  20.6& 3.74&  0.949&  29.4& 212.4& 142 \\
$p=-0.4$& this work &286.5&  46.6&  22.0& 3.74&  0.962&  32.5& 208.2& 142\\
$p= 0.0$& this work &286.5&  51.4&  23.4& 3.74&  0.974&  35.5& 204.1& 142\\
$p=+ 0.4$& this work &286.5&  56.2&  24.7& 3.74&  0.986&  38.6& 199.9& 142\\
$p=+ 0.8$& this work &286.5&  61.0&  26.1& 3.74&  0.999&  41.7& 195.7& 142\\
\hline
$o$ Draconids & 1 \citet{Denning1879a,Denning1879b} &    291 &  60 \\
$\theta$ Cygnids & 2 \citet{Denning1893} & 292 &   53 \\
$\alpha$ Lyrids &  3 \citet{Davidson1914} &  280 &   44 \\  
$\kappa$ Cygnids & 4 \citet{Whipple1954}\tablefootmark{a} &    288.2& 52.4& 24.2& 4.05&  0.975&  37.0& 203.4& 144.8 \\
$\kappa$ Cygnids &  5 \citet{Cook} & 286 &  59 &  24.8 &3.09&  0.99&   38&   194&   145 \\
$\kappa$ Cygnids & 6 \citet{Sekanina1973} &  299 & 62.4 & 25.9& 2.58&  0.979&  42.9 &203.1& 152.5 \\ 
$\kappa$ Cygnids & 8 \citet{Porubcan}\tablefootmark{b} &  278.4& 52.3& 21.7& 3.51&  0.991&  33.0 &197.7& 142.0 \\
$\kappa$ Cygnids & 9 \citet{Lindblad1995}  & 286.4& 55.1& 23.7 &3.72&  0.983&  36.8& 200.5& 145.0 \\
$\alpha$ Lyrids & \ \ \ ``&   283.7& 45.1& 21.8& 3.89&  0.961&  31.8& 208.0& 136 \\
$\kappa$ Cygnids & 10 \citet{Jones2006} & &&&  3.82&  0.981&  36.8& 201.3& 145.8 \\
$\alpha$ Lyrids &  \ \ \ \ \  ``&  &&&  3.76&  0.960&  31.5& 208.7& 136.2 \\
$\kappa$ Cygnids & 11 \citet{Trigo2009} &   283.6& 49.2& 22.5& 4.2&   0.974 & 33.4 &203.3& 142.1 \\
$\kappa$ Cygnids & 12 \citet{Jopek2003} &   284.3& 51.5& 23.2& 3.87&  0.975&  35.1& 202.9& 141.1 \\
12 KCG & 13 \citet{JennBook1}\tablefootmark{c} & 287.1& 49.5& 23.4& 4.10 & 0.957&  34.7 &206.2 &140.4\\
$\kappa$ Cygnids &  14 \citet{SonotaCo} &  285.0 & 50.1 & 21.9 \\
12 KCG & 15 \citet{Molau2009} &  285.9 & 51.0 & 22.7 \\
12 KCG (e)\tablefootmark{d} & 16 \citet{Molau2012} &  280.3& 50.1& 22.0\\
Group B & 17 \citet{Koseki2014}  & 286.8& 45.5& 21.3& 3.36&  0.958& 31.2& 209.1& 141.8 \\
Group C & \ \ \ \ \ \ ``&  286.8& 53.1& 23.2& 3.32&  0.977& 36.0& 203.6 &141.9  \\
Group D & \ \ \ \ \ \ ``&  290.1& 63.6& 27.2& 3.55  &0.993& 44.0& 195.4& 149.2 \\
12 KCG  & 23 \citet{Shiba2022} &  286.1& 49.1& 22.6& 3.52 & 0.968& 34.0& 206.2& 140.4 \\ 
12 KCG & 24 \citet{JennBook2} &  285.6& 51.4& 22.7& 3.34&  0.976& 34.7& 204.0& 140.8 \\ \hline
\end{tabular} \\
\tablefoottext{a}{average of his five meteors;}
\tablefoottext{b}{sample probably contamined by August Draconids;}
\tablefoottext{c}{DMS 1993 data with the lowest dispersion;}
\tablefoottext{d}{the early activity at $\lambda_\sun<142\degr$}
\end{table*}

\section{Refined parameters of the showers}
\label{refinement}

\subsection{$\kappa$ Cygnids}

Based on the comparison with previous work, Group~1 was identified with the $\kappa$ Cygnid meteor shower (Sect.~\ref{identification}). 
Here we present refined parameters of that shower based on our data.
The usual meteor shower parameters are the mean radiant, radiant daily motion, geocentric velocity and heliocentric orbit.
However, it is not possible to give such a simple set of parameters for $\kappa$ Cygnids. On one hand, there is a nice
dependency of right ascension on solar longitude (Fig.~\ref{solong-RA}). Two of the three most deviating meteors were observed outside the
year 2021 (in 2016 and 2018). On the other hand, there is no such dependency for declination (Fig.~\ref{solong-Dec}). 
While the mean declination increases by about 15\degr\
in 20 days, there is also a random scatter of about $\pm12\degr$ at any given time. To describe the deviation of declination from the mean value,
we introduced a variable $p \in\langle-1,1\rangle$. The value $p=0$ corresponds to the mean value, while the values  $p=\pm1$ are for the extreme
negative and positive deviations. Depending on the solar longitude $\lambda_\sun \in\langle130\degr,154\degr\rangle$ and $p$, the positions
of the $\kappa$ Cygnid radiant can be then written as
\begin{eqnarray}
\alpha_{\rm g} & = & 286\fdg5 + 0.45\ (\lambda_\sun -142\degr)\\
\delta_{\rm g} & = & +51\fdg4 + 0.73\ (\lambda_\sun -142\degr)+ 12\degr p,
\label{eqdelta}
\end{eqnarray}
where $\alpha_{\rm g}$ and $\delta_{\rm g}$ are the geocentric right ascension and declination, respectively.
 The solar longitude 142\degr\ ($\sim$ August 15) correspond to the middle (and the maximum) of the activity.

The variable $p$ can describe not only the variations in declination but also in other parameters. There is a pronounced correlation
between declination and velocity (Fig.~\ref{Dec-Vg}). Perihelion distance, inclination, and argument of perihelion are also
correlated with declination. We first fitted these parameters as a function of solar longitude to obtain the expected mean values at a given time.
Then the deviations of actual parameters from the expected means were plotted for individual meteors as a function of $p$ computed 
from Eq.~(\ref{eqdelta}). Linear dependencies were found in all cases. The maximal deviations for $p=\pm1$ were then found.
Only the core meteors observed in 2021 and 2024 were used.

\begin{figure}
    \centering
    \includegraphics[width=0.8\linewidth]{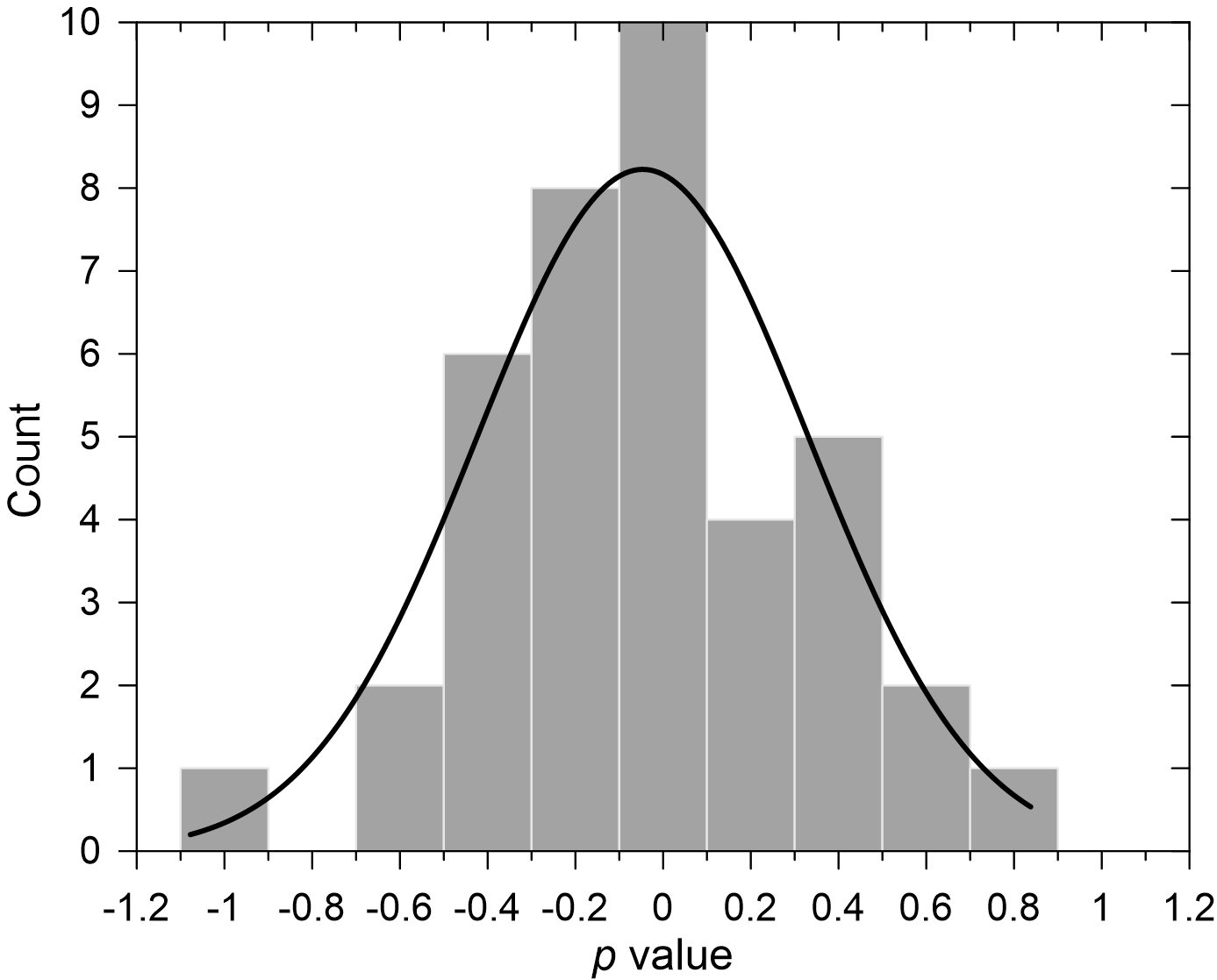}
    \caption{Histogram of values of $p$ variable (computed from Eq.~\ref{eqdelta}) for 39 members of Group~1 ($\kappa$ Cygnids).}
    \label{hist-p}
\end{figure}

The resulting equations for the velocity and orbit of the $\kappa$ Cygnid shower are:
\begin{eqnarray}
v_{\rm g} & = & 23.36 + 0.18\ (\lambda_\sun -142\degr)+ 3.4 p\\
a & = & 3.74 \pm 0.12  \\
q & = & 0.9739 + 0.00119\ (\lambda_\sun -142\degr)+ 0.031 p  \\
i & = & 35\fdg55 + 0.41\ (\lambda_\sun -142\degr)+ 7\fdg7 p \label{KCGi} \\
\omega & = & 204\fdg1 - 0.46\ (\lambda_\sun -142\degr) - 10\fdg4 p \\
\Omega &=& \lambda_\sun,
\end{eqnarray}
where $v_{\rm g}$ is geocentric velocity in km s$^{-1}$,  $a$ is semimajor axis, $q$ is perihelion distance (both in AU), $i$ is inclination, 
$\omega$ is argument of perihelion, and $\Omega$ is the longitude of ascending node (all in equinox J2000.0). 
The semimajor axis is the same for all meteors.
No trend was found. The given value is the average of 39 meteors. The 5:3 resonance is at 3.70~AU, i.e.\ well within the 
error bar. The value of $p$ for a given meteor can be computed from any of the equations which contains $p$. The result should
be similar (within about 0.2 in most cases). 
A nearly normal (Gaussian) distribution of $p$ values around zero was found (Fig.~\ref{hist-p}). In exceptional cases
the absolute value of $p$ can exceed unity.

Five representative radiants and orbits for selected values of $p$ are given in Table~\ref{KCGdata}. As it is customary, the orbits
are given for the maximum of the activity. They differ
just by $p$. While the dependence on $\lambda_\sun$ expresses the changes along the Earth's orbit, the dependence on $p$
describes the dispersion in perpendicular direction. Note that different orbits than given here may exist in the meteoroid stream. 
The shower data contain only orbits that intersect the Earth orbit.

Table~\ref{KCGdata} contains also detailed data on showers from the literature, which could be identified with $\kappa$ Cygnids 
(Sect.~\ref{identification}). Note that authors which performed a computer search in the orbital section of the IAU MDC
\citep{Porubcan,Lindblad1995,Jopek2003,Jones2006} partly used the same data, including those of \citet{Whipple1954}.

\subsection{August Draconids}

Based on the similarity of orbits, in particular the semimajor axis, and the absence of clear boundaries in radiant position
and velocity, Groups 2--4 can be considered as a single meteor shower. Following \citet{Shiba2017} and \citet{JennBook2},
we call the shower August Draconids (197/AUD). Nevertheless, the groups can be still recognized according to some characteristics.
They can be considered as three branches of August Draconids. Some showers, for example Taurids, have the northern and southern branch. 
Northern Taurids have the radiant north of ecliptic and meet the Earth at descending node. The 
opposite is true for Southern Taurids. Both branches meet the Earth on the way to perihelion. The perihelion point
of Northern Taurids lies below the ecliptic plane (to the south) and the perihelion point of Southern Taurids lies above the ecliptic.

In case of August Draconids, all radiants lie well north of ecliptic, in fact close to the pole of ecliptic. All branches meet the
Earth at descending node. As shown in Sect.~\ref{understanding}, Group~3 meets the Earth before reaching perihelion
and the perihelion lies below the ecliptic.  Group~4 meets the Earth after reaching perihelion
and the perihelion lies above the ecliptic.  Group~2 meets the Earth near perihelion, which lies close to the ecliptic. 
According to the location of the perihelion, we will call Group~3 the lower branch, Group~4 the upper branch, and Group~2 the
middle branch.

\begin{figure}
    \centering
    \includegraphics[width=0.9\linewidth]{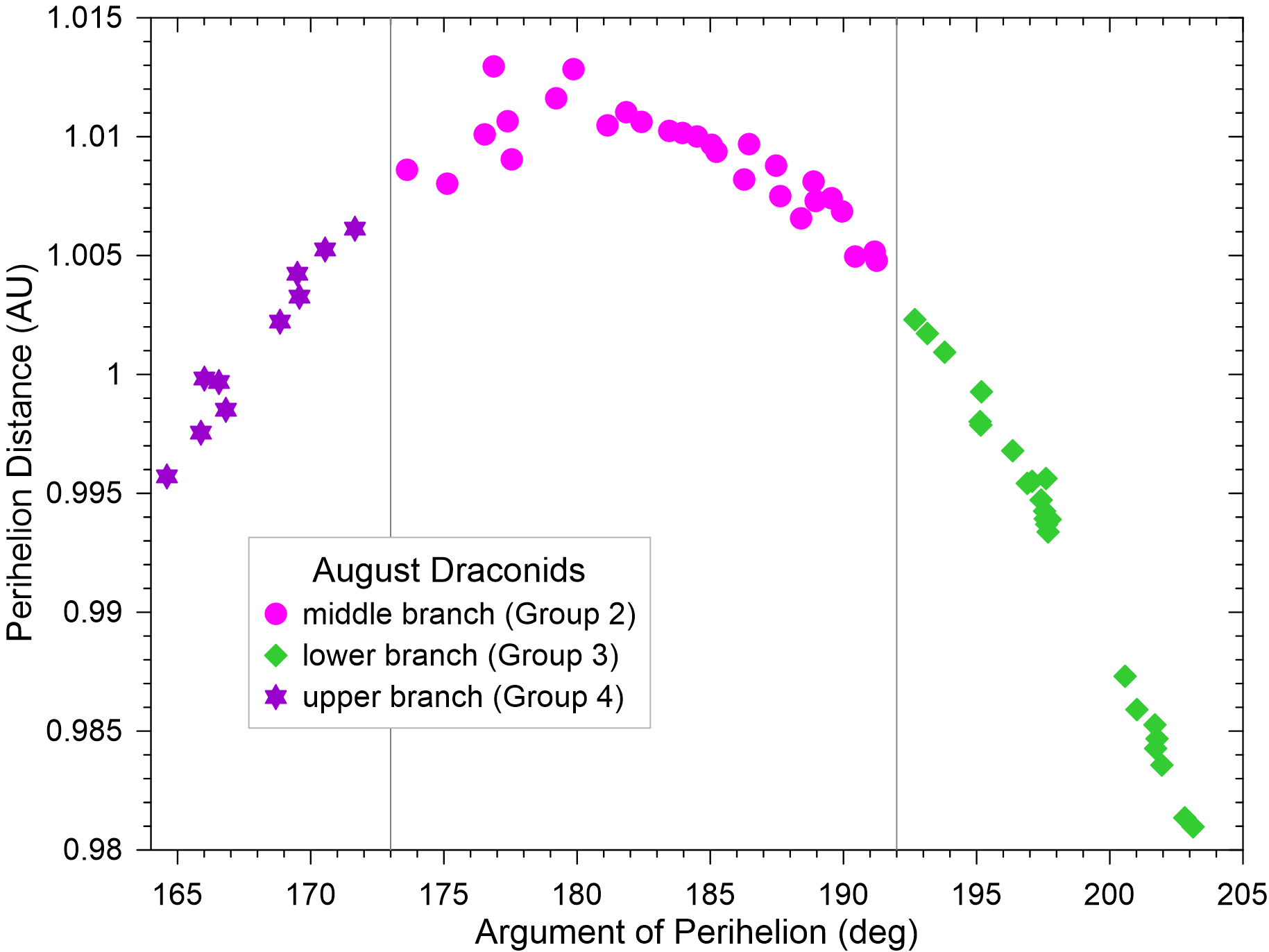}
    \caption{Relation between argument of perihelion and perihelion distance for three branches of August Draconids.}
    \label{argup-perih}
\end{figure}

The distinction between branches can be done either according to the radiant position (Fig.~\ref{RA-Dec})
or according to the argument of perihelion. The middle branch as defined by us has $\omega$ between 173\degr\ and 192\degr\
(Fig.~\ref{argup-i} and \ref{argup-perih}).

\subsubsection{Lower branch}

The radiant of the lower branch of August Draconids (Group~3) are tightly clustered in right ascension (Fig.~\ref{solong-RA})
but scattered in declination (Fig.~\ref{solong-Dec}). In both cases, an increase with solar longitude is observed.
The situation is similar to $\kappa$ Cygnids. In fact, the geometry of the orbits is also similar. The Earth encounter
occurs before the perihelion and the perihelion point lies to the south from the ecliptic. Similarly to $\kappa$ Cygnids,
the scatter in declination can be described by a variable $p$ reaching the values (approximately) from $-1$ to $+1$ with
normal distribution around zero. The scatter in geocentric velocity, perihelion distance, inclination, and argument of perihelion
can be also expressed by parameter $p$. We have found the following relations:
\begin{eqnarray}
\alpha_{\rm g} & = & 276\fdg1 + 0.20\ (\lambda_\sun -140\degr)\\
\delta_{\rm g} & = & +50\fdg6 + 0.60\ (\lambda_\sun -140\degr)+ 8\degr p \\
v_{\rm g} & = & 20.12 + 0.10\ (\lambda_\sun -140\degr)+ 2 p\\
a & = & 3.08 \pm 0.09  \\
q & = & 0.9926 + 0.00090\ (\lambda_\sun -140\degr)+ 0.012 p  \\
i & = & 30\fdg60 + 0.26\ (\lambda_\sun -140\degr)+ 4\fdg7 p \\
\omega & = & 198\fdg0 - 0.49\ (\lambda_\sun -140\degr) - 5\fdg3 p \\
\Omega &=& \lambda_\sun.
\end{eqnarray}
The members of this branch were observed in the interval of solar longitudes $\lambda_\sun$ from 130\degr to 148\degr.

The spread of the radiants and orbits as a function of $p$ is lower than in $\kappa$ Cygnids. We therefore give only three representative 
sets in Table~\ref{G3data}. The showers from the literature which can be identified with this branch are also listed there.

\begin{table*}
\caption{Three representative sets of radiants and orbits of the lower branch of August Draconids (Group~3)
and identification of corresponding showers from the literature.}
\label{G3data}
\begin{tabular}{llllllllll} \hline
Designation & Reference & $\alpha_{\rm g}$ & $\delta_{\rm g}$ &  $v_{\rm g}$ &  $a$ & $q$ & $i$ & $\omega$ & $\Omega$ \\
\hline
$p=-0.6$& this work &276.1 &  45.8 & 18.9 & 3.08&  0.985 & 27.8 &201.2 &140 \\
$p= 0.0$& this work &276.1 & 50.6  &20.1 &3.08 & 0.993  &30.6 &198.0 &140 \\
$p=+0.6$& this work &276.1 & 55.4  &21.3 &3.08 & 1.000&  33.4& 194.8 &140 \\
\hline
August Lyrids & 9  \citet{Lindblad1995} &  277.6 & 46.2 & 19.0 & 3.10 &  0.984 &  28.0 & 201.5 & 139.5 \\
August Lyrids & 10 \citet{Jones2006}  &&&&                               3.06 & 0.983 & 28.0& 201.6 & 140.2 \\
$\gamma$ Draconids&\ \ \ \ \ \ ``&  276.1 & 55.6  &&                           3.12  & 1.000 & 33.6& 193.9 & 142.2 \\
Group E & 17 \citet{Koseki2014} &  275.5 &44.7& 18.1& 2.77&  0.983& 26.5& 201.5& 139.9 \\
12 KCG & 18 \citet{CAMS} &    277.5 &52.8& 20.9& 2.95 & 0.995& 32.5& 196.9 & 140.0  \\
\hline
\end{tabular} \\
\end{table*}

\subsubsection{Middle branch}

The radiant of the  middle branch of August Draconids (Group 2) does not show any clear dependence on solar longitude.
The spread in both right ascension and declination is larger than any possible functional dependence  (Fig.~\ref{solong-RA}
and \ref{solong-Dec}). In fact, the radiants fill an irregular area in the vicinity of the north ecliptic pole (Fig.~\ref{RA-Dec}).
There is no concentration and no Gaussian distribution. Instead, the radiants cover the area more or less evenly. There
is a clear dependence of geocentric velocity, and also some orbital elements, on the position of the radiant.

To describe the dependencies quantitatively, we define the mean radiant position $\alpha_0,\delta_0$ and the deviation
from the mean position $\Delta\alpha,\Delta\delta$. With $\alpha_0=268\fdg4$ and $\delta_0=62\fdg0$, the actual
geocentric radiant is
\begin{eqnarray}
\alpha_{\rm g} & = & 268\fdg4 + \Delta\alpha\\
\delta_{\rm g} & = & +62\fdg0 + \Delta\delta,
\end{eqnarray}
from where $\Delta\alpha,\Delta\delta$ can be computed for any observed meteor. The observed range was 
$\Delta\alpha \in \langle -15\degr,+7\degr \rangle$ and 
$\Delta\delta \in \langle -5\degr,+5\degr \rangle$. The observed range of solar longitudes was
$\lambda_\sun \in \langle 138\degr,158\degr \rangle$.

The expected orbital elements for radiants and solar longitude within these ranges are:
\begin{eqnarray}
v_{\rm g} & = & 22.32 + \Delta\alpha/10 + \Delta\delta/2.6\\
a & = & 2.98 \pm 0.12  \\
q & = & 1.009 + 5\times 10^{-4} \Delta\delta  \\
i & = & 35\fdg67 + \Delta\alpha/5.2 + 0.75 \Delta\delta \\
\omega & = & 183\fdg9 + 0.64 \Delta\alpha - 0.7 \Delta\delta \\
\Omega &=& \lambda_\sun,
\end{eqnarray}
where $\Delta\alpha$ and $\Delta\delta$ are given in degrees.

\begin{table*}
\caption{Four representative sets of radiants and orbits of the middle branch of August Draconids (Group~2)
and identification of corresponding showers from the literature.}
\label{G2data}
\begin{tabular}{llllllllll} \hline
Designation & Reference & $\alpha_{\rm g}$ & $\delta_{\rm g}$ &  $v_{\rm g}$ &  $a$ & $q$ & $i$ & $\omega$ & $\Omega$ \\
\hline
$\Delta\alpha=+5, \Delta\delta=-4$& this work &  273.4&  58.0&  21.3& 2.98&  1.007&  33.7& 189.9& 148\\
$\Delta\alpha=0, \Delta\delta=0$& this work &  268.4 & 62.0 &  22.3& 2.98&  1.009&  35.7& 183.9& 148 \\
$\Delta\alpha=-4, \Delta\delta=+4$& this work &  264.4 & 66.0 & 23.5& 2.98&  1.011 & 37.9& 178.5& 148 \\
$\Delta\alpha=-10, \Delta\delta=0$& this work & 258.4 &  62.0 & 21.3& 2.98&  1.009&  33.8& 177.5 & 148 \\
\hline
August Draconids & 7 \citet{Sekanina1976} &  271.3& 65.3& 23.6& 2.78&  1.010&  38.5& 183.1& 140.8 \\
$\zeta$ Draconids & 9 \citet{Lindblad1995} &   267.9& 61.7& 22.0& 3.02&  1.007&  34.9& 181.7& 147.8\\
$\zeta$ Draconids &  10 \citet{Jones2006}  &&&&    2.91&  1.008&  34.5& 177.8& 152.0  \\
12 KCG (l)\tablefootmark{a} & 16 \citet{Molau2012} &  267.3& 61.6& 23.1\\
Group F & 17 \citet{Koseki2014} &   274.1& 57.3& 21.0& 2.74&  1.003& 33.3& 191.6& 143.9 \\
Group G & \ \ \ \ \ \ ``& 260.6& 63.0& 21.8& 2.83&  1.010& 34.8& 178.2& 149.7 \\
197 AUD &18 \citet{CAMS}&   271.7& 58.9& 21.1& 2.82&  1.008& 33.8& 188.7& 142.6 \\
470 AMD & 18 \citet{CAMS},&  256.4& 62.5& 21.3& 2.87&  1.009& 33.8& 175.5& 149.5 \\
 & 20 \citet{CAMS3} \\
470 AMD & 21 \citet{Rudawska}&    253.7& 58.8& 19.5& 2.92&  1.011& 30.3& 177.2& 145.4\\
470 AMD & 22 \citet{Kornos}& 254.8& 58.2& 19.0& 2.74&  1.012& 29.5& 178.4& 144.1 \\
197 AUD  & 23 \citet{Shiba2022}\tablefootmark{b} &&&  21.0 & 3.01&  1.001& 32.8& 184.6& 149.0 \\
197 AUD& 24 \citet{JennBook2}\tablefootmark{b} & 269.8 &59.3& 21.4& 2.93&  1.003& 33.6& 185.0& 148.5 \\
\hline
\end{tabular} \\
\tablefoottext{a}{the late activity at $\lambda_\sun>142\degr$;}
\tablefoottext{b}{combination of all three branches}
\end{table*}

Four representative sets of radiants and orbits are given
in Table~\ref{G2data}. The showers from the literature which can be identified with this branch are also listed.
Not listed are $\xi$ Draconids \citep[][listed under 220/NDR in the IAU MDC]{Sekanina1976} and 
$\nu$ Draconids \citep[220/NDR,][]{CAMS2}. These proposed
showers have similar radiants and orbits but their period of activity is at a later time (and the longitude of the ascending node
is therefore larger) than it is the case for the middle branch of August Draconids.

\subsubsection{Upper branch}

The number of observed meteors in the upper branch (Group~4) was lower than in other two branches and in $\kappa$ Cygnids.
Since no dependency of radiant position on solar longitude is apparent (Figs.~\ref{solong-RA}, \ref{solong-Dec}), we used
the same approach as for the middle branch to describe the extend of orbital elements. The following equations were found:
\begin{eqnarray}
\alpha_{\rm g} & = & 242\fdg9 + \Delta\alpha\\
\delta_{\rm g} & = & +60\fdg3 + \Delta\delta \\
v_{\rm g} & = & 20.40 + \Delta\delta/3\\
a & = & 3.06 \pm 0.08  \\
q & = & 1.0012 + 7.5\times 10^{-4} \Delta\alpha  \\
i & = & 31\fdg3 + 0.75 \Delta\delta \\
\omega & = & 168\fdg0 + 0.55 \Delta\alpha \\
\Omega &=& \lambda_\sun,
\end{eqnarray}
with observed limits 
$\Delta\alpha \in \langle -6\degr,+6\degr \rangle$, 
$\Delta\delta \in \langle -5\degr,+2\degr \rangle$, and
$\lambda_\sun \in \langle 148\degr,160\degr \rangle$.

Three representative sets of radiants and orbits are given
in Table~\ref{G4data}. The only shower from the literature which could be identified with the upper branch are $\iota$ Draconids.
They do not correspond exactly with the branch parameters but lie in the prolongation towards lower $\omega$.

\begin{table*}
\caption{Three representative sets of radiants and orbits of the upper branch of August Draconids (Group~4)
and identification of a corresponding shower from the literature.}
\label{G4data}
\begin{tabular}{llllllllll} \hline
Designation & Reference & $\alpha_{\rm g}$ & $\delta_{\rm g}$ &  $v_{\rm g}$ &  $a$ & $q$ & $i$ & $\omega$ & $\Omega$ \\
\hline
$\Delta\alpha=+5, \Delta\delta=-1$& this work &247.9&  61.3&  20.7& 3.06 & 1.005&  32.0& 170.8& 154 \\
$\Delta\alpha=0, \Delta\delta=0$& this work & 242.9 & 60.3&  20.4& 3.06&  1.001&  31.3& 168.0& 154 \\
$\Delta\alpha=-6, \Delta\delta=-4$& this work &  236.9& 56.3&  19.1& 3.06&  0.997 & 28.3& 164.7& 154\\
\hline
703 IOD & 18 \citet{CAMS} &    232.3& 53.3& 17.8& 2.93&  0.990& 26.1& 161.5 &157.2 \\
\hline
\end{tabular} \\
\end{table*}

 \begin{figure}
    \centering
    \includegraphics[width=0.9\linewidth]{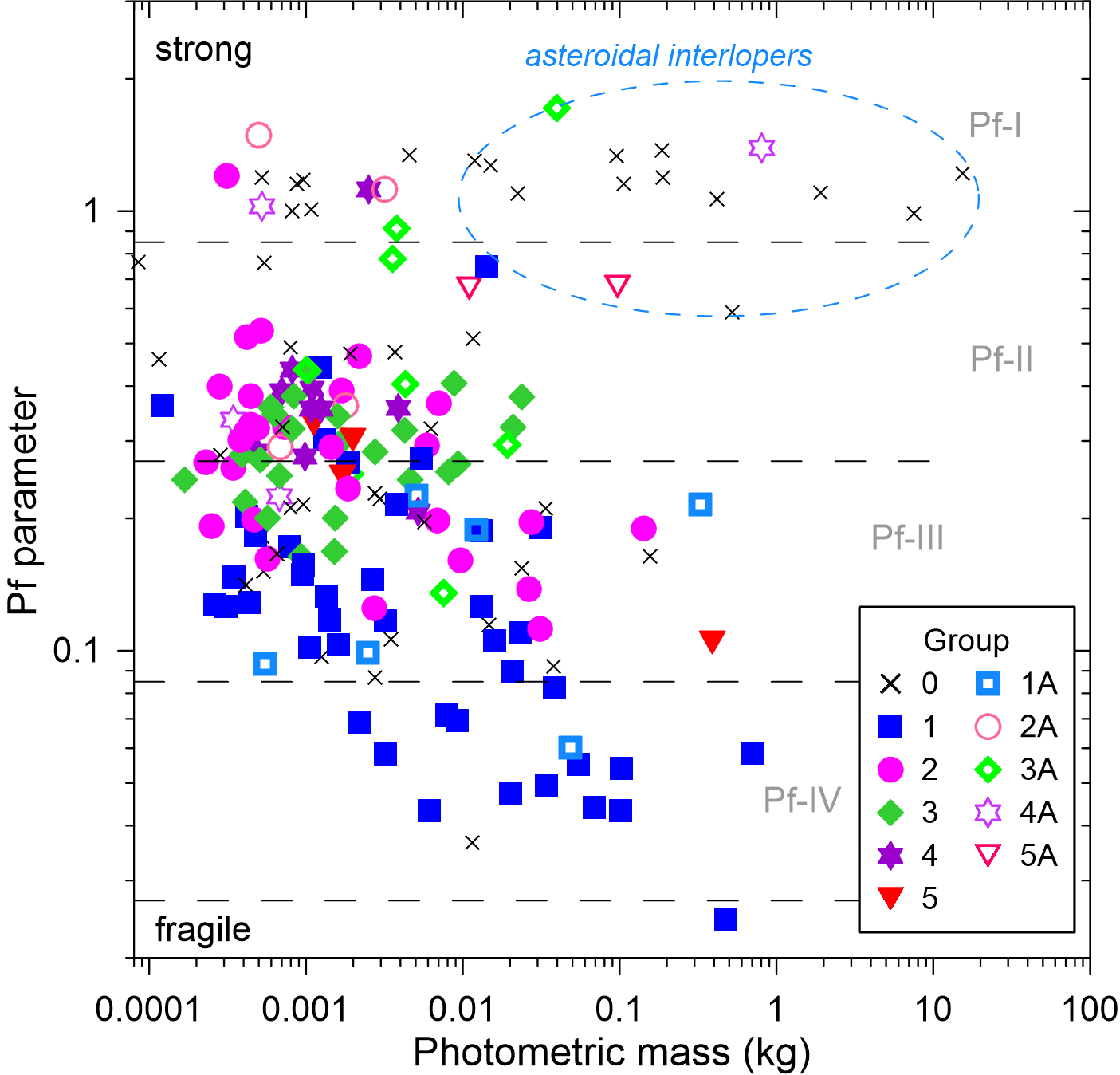}
    \caption{Pressure factor $P\!f$ describing meteoroid mechanical strength as a function of mass for all studied meteors.
    Orbital groups as defined in Figs.~\ref{radiants-polar} and \ref{periods} are marked by different symbols.
    The boundaries between Pf-classes as defined by \citet{catalog2} are also shown.}
    \label{mass-Pf}
\end{figure}

\section{Physical classification}
\label{physical}

The basic physical classification of meteoroids can be done using the PE criterion
of \citet{PE} or using the Pressure-factor ($P\!f$) recently defined by \citet{catalog2}.
The PE criterion is based for the fireball terminal height for a given mass and velocity
while $P\!f$ considers the maximal dynamic pressure. Both approaches gave the same 
picture in the present case. 

Figure~\ref{mass-Pf} shows $P\!f$ as a function of
meteoroid mass for all studied meteors. A part of the meteoroids was strong with $P\!f\sim1$.
That value correspond to material similar to stony meteorites. Most of these meteoroids were sporadic. 
Some of them were suspected shower members based on their radiants but not confirmed according
to their orbital period (plotted with open symbols).  Only three meteors were classified as shower
members on the basis of the orbital elements. Two as August Draconids and one as a $\kappa$ Cygnid.
All of them were small with masses $\la 10$ grams. Considering the large width of the streams, we
cannot exclude that these meteoroids were also interlopers from sporadic background. 
Alternatively, they could represent a minority strong fraction of the shower material.
In any case, they exclusion from the analysis would not affect the results of the previous sections.

A large majority of shower meteors had $P\!f<0.5$. There is a clear trend of decreasing strength
with increasing mass. This effect was previously observed in Taurids \citep{Taurids}.  
It has been found that large bodies are composed almost entirely of weak material, which contains stronger inclusions.
Small meteoroids represent these inclusions \citep{Taurids2}.
It seems that $\kappa$ Cygnids and August Draconids have similarly hierarchical structure as Taurids. Such a structure is
probably usual for cometary material on Jupiter-family orbits. 
At the same time, it is evident that $\kappa$ Cygnids are
on average significantly more fragile than August Draconids. That fact confirms that they are 
different showers with different origin. $\kappa$ Cygnids also contain larger fraction of more massive
meteoroids. Masses up to 1 kg were observed. There is no difference in physical properties of individual branches 
of August Draconids.

A more detailed study of physical properties including fragmentation modeling such as done for
Taurids \citep{Taurids2} is left for future work.

\section{Discussion}

A meteor shower activity in August with radiants in the Cygnus-Draco region has been studied for
almost 150 years. The radiant situation seemed to be very complex. More recent work
suggested that only two showers are in action, $\kappa$ Cygnids and August Draconids \citep{Shiba2017, JennBook2}. Our
work confirmed this conclusion. Both showers have, nevertheless, a complex structure
and cannot be described by a single set of radiant coordinates and orbital elements. We have
provided equations to compute the plausible range of orbital elements and listed sets of typical
elements within that ranges. The ranges cover the elements that can be currently registered on Earth.
We cannot exclude that different combinations of orbital elements, which do not lead
to Earth intersection, exist within the streams.

\begin{table}
\caption{Years of expected activity of $\kappa$ Cygnids}
\label{years}
\raisebox{-0.7in}{\rotatebox{90}{$<$--- 7 year interval ---$>$}}
\hspace{0.1in}
\begin{tabular}{llll}
\multicolumn{4}{c}{$<$--- 64 year interval ---$>$} \\ \hline
1865& 1929&\bf 1993& 2057 \\
1872& 1936& 2000& 2064\\
\bf 1879& 1943&\bf 2007&2071 \\
1886&\bf 1950&\bf 2014& 2078\\
\bf 1893& 1957&\bf 2021& 2085\\
1900& 1964& 2028&\\
\bf 1907& 1971& 2035&\\
\bf 1914&\bf 1978& 2042&\\
\bf 1921&\bf 1985& 2049&\\ \hline
\end{tabular}
\end{table}

\subsection{Periodicity}
\label{periodicity_discus}

One important topic discussed in the past is the 7-year periodicity of $\kappa$ Cygnids and
their presence in a  5:3 resonant swarm with Jupiter. We were able to confirm that hypothesis by direct measurements
of meteoroid orbital periods. They really cluster around the expected resonant value of 7.117 years. 
The measurements are still not precise enough to show that the periods sit exactly at this value but considering long
term periodicity it is almost certain. Since the period is larger than 7 years, the activity is not expected to repeat
identically after 7 years. Nine cycles take 64.05 years, so the situation will repeat closely after 64 years.
The evolution in between depends on how are the meteoroids distributed in mean anomaly.

Table~\ref{years} shows the expected years of activity.
Our data showed that there was high activity in 2021 but no noticeable activity in the neighboring years 2020 and 2022,
when the stream was sampled at $\pm 50\degr$ in mean anomaly from the maximum in 2021.
The extent of the swarm is therefore lower than 100\degr.
The next maximum is expected in 2028 and then every 7 years until 2049. 
But the maximum will shift forward in time and is expected to be stronger in 2057 than in 2056. In fact, there are two
possibilities. If the swarm is very narrow in mean anomaly, the maximum can fall in between 2049 and 2050 
and then in between 2056 and 2057. No activity may be
observed in any of these years. But if the swarm extent is larger, activity can occur in all these years at a similar level,
lower than in 2021.

To decide between these two possibilities, we can look into the past. The years when a $\kappa$ Cygnids outburst was reported in the literature
(see Sect.~\ref{history}) are marked in bold in Table~\ref{years}. 
In fact, all well documented outbursts occurred in years listed in Table~\ref{years}.
These years include 1985 and 1993, which are equivalent to 2049 and 2057, respectively, when the 64 year period is considered. Since the
activity was detected in these transition years (8 years apart), the swarm cannot be very narrow.

In some years, no enhanced activity was reported though it could be expected. In more distant past, that fact can be ascribed 
to lack of observations. The most suspicious is the year 2000. \citet{SPA2000} reported low visual activity of $\kappa$ Cygnids in 2000
but noted that the expected peak was lost to moonlight (the full Moon occurred on August 15, 2000). 
It is interesting that \citet{Miskotte2020} mentioned
an outburst in 1999, but without any detail. \citet{SPA1999} reported that though $\kappa$ Cygnids were observed
throughout August 1999, the level of activity was unimpressive. Checking the archive photographs from the European Fireball Network, 
we did not find any high $\kappa$ Cygnid fireball activity neither in 1999 nor in 2000.

\begin{figure}
    \centering
    \includegraphics[width=0.9\linewidth]{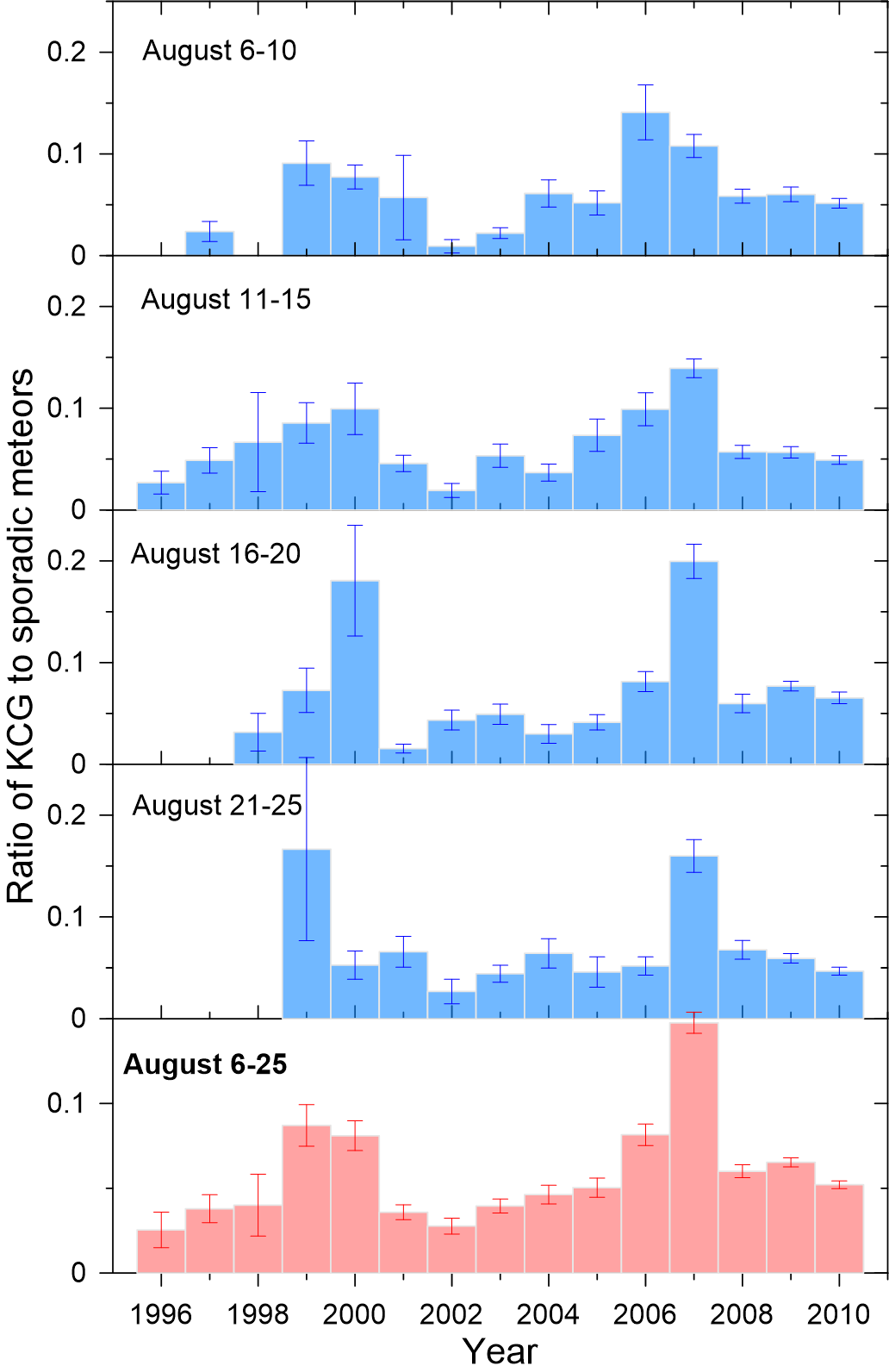}
    \caption{Ratio of meteors classified as $\kappa$ Cygnids to sporadic meteors as observed by the IMO Video Meteor Network in 1996--2010.
    The upper four panels show the data for five-day intervals. The lower panel shows the summary data for the whole studied
    period August 6-26. The statistical uncertainties are indicated.}
    \label{VMNdata}
\end{figure}

Trying to find more data, we checked the archive\footnote{https://meteorflux.org/obs\#selectiontab}
of the International Meteor Organization (IMO) Video Meteor 
Network \citep{Molau_VMN}. The archive contains meteor shower assignments 
based on single station video observations. To obtain information about the activity level of $\kappa$ Cygnids, the ratio of
the number of observed $\kappa$ Cygnids to the number of observed sporadic meteors was computed in four five-day intervals between
August 6--25 in 1996--2010 (Fig.~\ref{VMNdata}). The data from 1996--1998 are sparse since the number of cameras was low.
 We must also keep in mind that meteors classified as $\kappa$ Cygnids include very
probably also August Draconids, at least the lower branch. 

The high activity in 2007 is clearly visible in Fig.~\ref{VMNdata}. A higher activity than normal was observed also in 1999, 2000, and 2006. 
The activity in 2000, and thus in all expected years from Table~\ref{years} since 1978, can be therefore confirmed. 
Moreover, the activity was observed also in the years 1999 (probably at similar level as in 2000), 2006 (lower than in 2007), and,
as \citet{Shiba2017} reported, in 2013 (much lower than in 2014). This fact is not surprising and suggests that the width of the resonant swarm 
approaches $\sim90\degr$ in mean anomaly. We can therefore expect the activity to occur also in the preceding years to those listed in the upper part
of Table~\ref{years} and in the following years to those listed in the lower part. The years in the middle row, such as 2021, should show a high
activity with no activity in neighboring years. In this view, it is somewhat surprising that no activity was reported in 1986 and in 1992.
But we must realize that the high activity is never very high in terms of ZHR and represents just several meteors seen visually per hour.
What is remarkable is the presence of fireballs and that the activity lasts for at least several nights.

The situation may be more complex owing to the fact that the $\kappa$ Cygnid stream is extended in longitude of the node and in
inclination. According to \citet{Shiba2017}, the 2013 activity occurred at low solar longitudes (in early August) 
and contained orbits with high inclinations. Our analysis showed that the inclination generally increased
with solar longitude in 2021 (see Eq.~\ref{KCGi}).  Different parts of the swarm can therefore have different distributions in solar longitude
and inclination. The 2006 activity also
occurred in early August while the 1999 activity seems to predominate at later times (Fig.~\ref{VMNdata}). More data from different
years are needed to study these details.

August Draconids do not show such a pronounced periodicity as $\kappa$ Cygnids. Nevertheless,
a minor variation from year to year cannot be excluded.

\subsection{Parent body}

Several authors tried to find the parent body of $\kappa$ Cygnids. \citet{Jones2006} noted that the parent body should
not only have a similar orbit but should also exhibit similar orbital evolution as the stream meteoroids when integrated backwards.
They found that their substreams August Lyrids, $\gamma$ Draconids, and $\zeta$ Draconids, which are all part of August Draconids (197/AUD)
in our view, show sinusoidal variations of some orbital elements (namely eccentricity, perihelion distance, inclination) 
with a period of the order of 2000 years.
In contrast, $\kappa$ Cygnids and $\alpha$ Lyrids, which are also part of $\kappa$ Cygnids (12/KCG) in our view, showed flat behavior, in some cases
with quick oscillations. \citet{Jones2006} proposed asteroids 2001 MG1 and 2004 LA12 as possible parent bodies of $\kappa$ Cygnids.

\citet{JennVaub} found that the newly discovered asteroid 2008 ED69 shows similar sinusoidal variations in orbital evolution as described
by \citet{Jones2006} and proposed 2008 ED69 as the parent body of $\kappa$ Cygnids. However, that behavior corresponds to August Draconids.
Indeed, \citet{JennBook2} lists 2008 ED69 as the probable parent body of August Draconids.

\citet{Moorhead} simulated the transport of meteoroids from several asteroids and found again 2001 MG1 to be the best candidate though the
match of simulated radiants and velocities with $\kappa$ Cygnids was far from perfect.  \citet{JennBook2} proposes asteroid 2021 HK12 as the
source of   $\kappa$ Cygnids without further details.

\begin{table}
\caption{Orbital elements of previously proposed parent bodies of $\kappa$~Cygnids (upper part) and objects
with similar ($a,q,i$) as $\kappa$ Cygnids (lower part)}
\label{parents} 
\begin{tabular}{llllllll} \hline
Object & $a$ & $q$ & $i$ & $\omega$ & $\Omega$ & $\pi$ & $H$ \\ 
& AU & AU & \degr & \degr & \degr &  \degr &  mag \\ \hline
153311 \\[-0.2ex]
(2001 MG1) & 2.50 & 0.89 & 28 & 219& 142 & 1 & 17.4 \\
144861 \\[-0.2ex]
(2004 LA12) & 2.51 & 0.63 & 39 & 199& 159 & 359 & 15.4 \\
361861 \\[-0.2ex]
(2008 ED69) & 2.91 & 0.76 & 36 & 173& 150 & 323 & 17.0 \\
(2021 HK12) & 3.29 & 1.04 & 47 & 169& 187 & 356 & 17.7 \\ \hline
 && 0.94&28 & 190 & 130 & 335 \\[-0.2ex]
$\kappa$ Cygnids & 3.7 & 1.01 & 44 & 215 & 154 & 355 \\ \hline
(2001 XQ) &	3.63&	1.01&	29&	190&	251&80&		19.2	\\
(2010 HY22)&	3.66&	1.01&	44&	343&	19&2&	16.7\\
(2015 OA22)&	3.51&	1.07&	45&	357&	347&344&	19.8\\	
21P/Giacobini	\\[-0.2ex]	
-Zinner&	3.5&	1.01&	32&	173&	195&8 \\ \hline		
\end{tabular}
\small{Semimajor axis, perihelion distance, inclination, argument of perihelion, longitude of the
ascending node, longitude of the perihelion, and absolute magnitude are given. Upper and lower limits
are given for $\kappa$~Cygnids.}
\end{table}

The orbital elements of the proposed parent bodies are given in Table~\ref{parents}. All of them have semimajor axis
significantly smaller than $\kappa$ Cygnids. We have searched the small body database\footnote{https://ssd.jpl.nasa.gov/}
for bodies having the semimajor axis, $a$, perihelion distance, $q$, and inclination, $i$, similar to $\kappa$ Cygnids.
Four objects were found within the limits $3.5<a<3.9$ AU, $0.9<q<1.1$ AU, and $25<i<50\degr$. They
are listed in the lower part of Table~\ref{parents}. Three of them are asteroids, the fourth is comet 21P/Giacobini-Zinner,
the parent comet of the October Draconid meteor shower. Little is known about the three asteroids.
2015 OA22 has a low albedo (1.3\%), suggesting
a primitive composition. The albedos of the remaining two asteroids are not known. In any case, the angular elements 
($\omega, \Omega$) of all four objects are markedly different from $\kappa$ Cygnids. It is, nevertheless, worth mentioning
that except 2001 XQ, the longitude of perihelia ($\pi=\Omega+\omega$) of the other three objects 
are not much different from $\kappa$ Cygnids.

Physical properties of $\kappa$ Cygnids are distinctly cometary. They belong to the weakest known meteoroids, which is also
true to October Draconids. Comet 21P was rejected as a possible parent body of $\kappa$ Cygnids by \citet{Jones2006} because
of the difference in the longitude of the node. More recently, \citet{Neslusan} studied the past orbital evolution of 21P and simulated the
associated meteoroid stream. They noted rapid orbital evolution. The comet orbit can be considered reliably known only during the last 1500 years.
The ejected meteoroids are expected to be dispersed into sporadic background within 1000 years. We can speculate that 
a part of the stream was trapped into the 5:3 resonance several thousands years ago, avoided further dispersion, 
and forms the $\kappa$ Cygnids today. But it will be difficult to prove such scenario. More work is also needed evaluate the
possible relation of asteroids from Table~\ref{parents}, in particular 2010 HY22 and 2015 OA22, to $\kappa$ Cygnids.

\section{Conclusions}

We have proposed a clarification of the situation with meteor showers active in August and having radiants in the Cygnus-Draco area.
There are two separate showers referred to as $\kappa$ Cygnids (12/KCG) and August Draconids (197/AUD). Both showers are extended systems
active (at low level) for most of the month and having large radiant areas and wide range of entry velocities. 
Their orbits cannot be described by a single set of orbital elements.

The more spectacular are $\kappa$ Cygnids which are rich in bright, often bursting, fireballs produced by extremely fragile
cometary meteoroids with masses up to at least one kilogram. A noticeable activity, however, does not occur every year. Most
of $\kappa$ Cygnid activity is produced by an isolated swarm of meteoroids locked in the 5:3 main-motion resonance with Jupiter.
The orbital period is 7.12 years and we estimate the extent of the swarm to be at most 90\degr in mean anomaly
($\pm 45\degr$ from the center). The shower can be therefore observed in a single year or, at most, during two consecutive years
(but at a lower level) within each 7-year period. Only occasional $\kappa$ Cygnids can appear in other years.

The activity mostly occurs at solar longitudes 130--152\degr. The radiant extends over almost 30\degr\ in declination.
Some authors considered more showers to be active within that area but we showed that it is a single shower. 
The extend in declination reflects the extent in inclinations of the orbits, which is 28--44\degr. A general
increase of mean inclination with solar longitude was observed in 2021, but meteoroids with wide range of inclinations ($\pm8\degr$)
could be encountered at any given time. The inclination is correlated not only with declination but also with geocentric velocity,
perihelion distance, and argument of perihelion. We defined a variable $p$, reaching values between $-1$ and $+1$,
to describe the correlated deviations of these parameters from their mean values. The reason for the correlation
is probably the condition that the orbit must intersect Earth's orbit.

The August Draconids are even more complex. The whole stream can be divided into three branches. The middle
branch has perihelia and aphelia near the nodes. The perihelia lie close to the Earth's orbit and radiants are close
to the northern pole of ecliptic. The lower branch has perihelia south of the ecliptic plane and radiants 
to the south of the ecliptic pole. The upper branch has perihelia north of the ecliptic plane and radiants 
to the west of the ecliptic pole. The activity period of the lower branch is similar to that of $\kappa$ Cygnids. Because
the radiants are not far apart, these two showers have sometimes been confused. Nevertheless, August Draconids
have lower velocities and shorter semimajor axes. The activity of the middle branch is shifted to later times and the
upper branch is active even later. Nevertheless, the activity periods of the branches partly overlap. It is possible that
the system is wider and the activity starts already in July and ends in September. It is a matter of future work to check
if other proposed showers such as $\kappa$ Lyrids (464/KLY) active in July or $\nu$ Draconids (220/NDR) active in September are part
of the August Draconid complex.

August Draconids seem to be active every year, though some variations in the intensity cannot be excluded. The
orbits are close to the 7:3, 9:4, and 11:5 resonances with Jupiter with periods between 5 -- 5.5 years. It is not clear, however,
how many meteoroids are indeed in the resonances.
The meteoroids are more compact than $\kappa$ Cygnids though still cometary. The stream also does not
contain so many large ($>50$~g) meteoroids as $\kappa$ Cygnids. The candidate parent body is asteroid 2008 ED69, though the match is not
assured in our opinion. The parent body of $\kappa$ Cygnids is even more elusive. Taking into account its chaotic orbit,
we do not exclude comet 21P/Giacobini-Zinner as the parent body, though the current orbit is different.

In summary, the precise data from the European Fireball Network enabled us to describe the orbital structure of
$\kappa$ Cygnids and August Draconids. The directions of future research can be the detailed modeling of atmospheric
fragmentation to reveal the material properties on one hand and orbital interactions to search for parent bodies on the other hand.

\section*{Data availability}

The trajectories, radiants, orbits, magnitudes, and physical classifications of all 179 fireballs used for the present analysis are
available at the \textit{Centre de Donn\'ees astronomiques de Strasbourg} (CDS). The format is almost identical to our
previous catalog of 824 fireballs \citep{catalog}. Eighteen fireballs are present in both catalogs.
Instead of shower assignment, we give here the group number as in Fig.~\ref{radiants-polar},
G1--G5 for the core members and G1A--G5A for possible members (deviating in orbital period) of the groups. 
G1 is identical with $\kappa$ Cygnids, G2--G4 are three branches of August Draconids. Sporadic fireballs
with the radiants in the region of interest are included as well.

\begin{acknowledgements}
We thank all technicians and operators of the EN for their work. Student Anna Marie Van\v{c}urov\'a
helped us with the measurement of some fireballs and the initial analysis.
We acknowledge the use of the IAU Meteor Data Center Shower Database.
We are grateful to an anonymous referee for carefully checking the manuscript and providing detailed comments.
This study was supported by grant no.\ 24-10143S from Czech Science Foundation. 
\end{acknowledgements}


\begin{thebibliography}{}

\bibitem[Besley(1901)]{Besley1901}
Besley, W. E.\ 1901, JBAA, 12, 60

\bibitem[Borovi{\v{c}}ka \& Spurn{\'y}(2020)]{Taurids2}
Borovi{\v{c}}ka, J., \& Spurn{\'y}, P.\ 2020, 
\planss, 182, 104849

\bibitem[Borovi{\v{c}}ka et al.(2022a)]{catalog}
Borovi{\v{c}}ka, J., Spurn{\'y}, P., Shrben{\'y}, L., et al.\ 2022a,
\aap, 667:A157

\bibitem[Borovi{\v{c}}ka et al.(2022b)]{catalog2}
Borovi{\v{c}}ka, J., Spurn{\'y}, P., and Shrben{\'y}, L., \ 2022b,
\aap, 667:A158

\bibitem[Ceplecha \& McCrosky(1976)]{PE}
Ceplecha, Z., \& McCrosky, R.~E. 1976,
JGR, 81, 6257 %--6275.

\bibitem[Cook(1924)]{Cook1924}
Cook, A. G.\ 1924, Memoirs BAA, 24, 49

\bibitem[Cook(1973)]{Cook}
Cook, A. F.\ 1973, In: Evolutionary and Physical Properties of Meteoroids, C. L. Hemenway, P. M. Millman,
\& A. F. Cook (eds.), NASA-SP 319, p. 183

\bibitem[Davidson(1914)]{Davidson1914}
Davidson, M.\ 1914, JBAA, 25, 125

\bibitem[Denning(1879a)]{Denning1879a}
Denning, W. F.\ 1879a, \nat, 20, 457

\bibitem[Denning(1879b)]{Denning1879b}
Denning, W. F.\ 1879b, Observatory, 3, 170

\bibitem[Denning(1893)]{Denning1893}
Denning, W. F.\ 1893, Observatory, 16, 317

\bibitem[Denning(1899)]{Denning1899}
Denning, W. F.\ 1899, Memoirs RAS, 16, 203

\bibitem[Denning(1907)]{Denning1907}
Denning, W. F.\ 1907, \nat, 76, 413

\bibitem[Hajdukov{\'a} et al.(2023)]{MDC} 
Hajdukov{\'a}, M., Rudawska, R., Jopek, T.~J., et al.\ 2023, \aap, 671, A155

\bibitem[Jenniskens(1994)]{Jenn1994} 
Jenniskens, P. 1994, \aap, 287, 990

\bibitem[Jenniskens(2006)]{JennBook1} 
Jenniskens, P. 2006, Meteor Showers and their Parent Comets (Cambridge:
Cambridge University Press)

\bibitem[Jenniskens(2021)]{CBET2021} 
Jenniskens, P.\ 2021, Central Bureau Electronic Telegram \#5014

\bibitem[Jenniskens(2023)]{JennBook2} 
Jenniskens, P. 2023, Atlas of Earth's Meteor Showers (Elsevier)

\bibitem[Jenniskens \& N\'enon(2016)]{CAMS3} 
Jenniskens, P., and N\'enon, Q. 2016, \icarus, 266, 371

\bibitem[Jenniskens \& Trigo-Rodriguez(2007)]{CBET2007} 
Jenniskens, P. \& Trigo-Rodriguez, J. M.\ 2007, Central Bureau Electronic Telegram \#1055

\bibitem[Jenniskens \& Vaubaillon(2008)]{JennVaub} 
Jenniskens, P., \& Vaubaillon, J. 2008, \aj, 136, 725

\bibitem[Jenniskens et al.(2016a)]{CAMS} 
Jenniskens, P., N\'enon, Q., Albers, J. et al.\ 2016a, \icarus, 266, 331

\bibitem[Jenniskens et al.(2016b)]{CAMS2} 
Jenniskens, P., N\'enon, Q., Gural, P. S. et al.\ 2016b, \icarus, 266, 355

\bibitem[Jones et al.(2006)]{Jones2006}
Jones, D. C., Williams, I. P., \& Porub\v{c}an, V.\ 2006, \mnras, 371, 684

\bibitem[Jopek et al.(2003)]{Jopek2003}
Jopek, T. J., Valsecchi, G. B., \& Froeschl\'e, Cl.\ 2003, \mnras, 344, 665

\bibitem[Jopek et al.(2024)]{Jopek2024}
Jopek T.~J., Neslu{\v{s}}an L., Rudawska R., \& Hajdukov{\'a} M.\ 2024, \aap, 682, A159

\bibitem[Korno\v{s} et al.(2014)]{Kornos}
Korno\v{s}, L., Matlovi\v{c}, P., Rudawska, R., et al.\ 2014, in Meteoroids 2013, T. J. Jopek et al. (eds.)
A. M. University Press, Pozna\'n, Poland, p. 225

\bibitem[Koseki(2014)]{Koseki2014}
Koseki, M.\ 2014, WGN, J. IMO,  42, 181

\bibitem[Koseki(2020)]{Koseki2020}
Koseki, M.\ 2020, WGN, J. IMO,  48, 130

\bibitem[Letfus(1955)]{Letfus}
Letfus, V.\ 1955, \bac, 6, 143

\bibitem[Lindblad(1995)]{Lindblad1995}
Lindblad, B.\ 1995, Earth, Moon \& Planets, 68, 397

\bibitem[McBeath(2000)]{SPA1999}
McBeath, A.\ 2000, WGN, J. IMO, 28, 84

\bibitem[McBeath(2001)]{SPA2000}
McBeath, A.\ 2001, WGN, J. IMO, 29, 55

\bibitem[Miskotte(2020)]{Miskotte2020}
Miskotte, K.\  2020, eMeteorNews, 5, 161

\bibitem[Miskotte et al.(2022)]{Miskotte2022}
Miskotte, K., Johannink, C. \& Betlem H.\ 2022, eMeteorNews, 7, 19

\bibitem[Molau \& Barentsen(2014)]{Molau_VMN}
Molau, S. \& Barentsen, G.\ 2014, in Meteoroids 2013, T. J. Jopek et al. (eds.)
A. M. University Press, Pozna\'n, Poland, p. 297

\bibitem[Molau \& Rendtel(2009)]{Molau2009}
 Molau, S.  \& Rendtel, J.\ 2009, WGN, J. IMO,  37, 98

\bibitem[Molau et al.(2012)]{Molau2012}
Molau, S., Kac, J, Berko, E.. et al.\ 2012, WGN, J. IMO,  40, 201

\bibitem[Molau et al.(2015)]{Molau2015}
Molau, S., Crivello, S., Goncalves, R. et al.\ 2015, WGN, J. IMO,  43, 188

\bibitem[Moorhead et al.(2015)]{Moorhead}
Moorhead, A. V., Brown, P. G., Spurn\'y, P., Cooke, W. J., \& Shrben\'y, L.\ 
2015, \aj, 150, 122

\bibitem[Neslu\v{s}an \& Tomko(2023)]{Neslusan}
Neslu\v{s}an, L. \& Tomko, D.\ 2023, \icarus, 392, 115375

\bibitem[Porub\v{c}an  \& Gavajdov\'a(1994)]{Porubcan}
Porub\v{c}an, V.  \& Gavajdov\'a. M.\ 1994, Planet. Space Sci., 42, 155

\bibitem[Prentice et al.(1929)]{Obs1929}
Prentice, J. P. M. et al.\ 1929, Observatory, 52, 310

\bibitem[Rendtel \& Arlt(2016)]{RendtelArlt}
Rendtel, J., \& Arlt, R.\ 2016, WGN, J. IMO,  44, 62

\bibitem[Rendtel \& Molau(2015)]{RendtelMolau}
Rendtel, J., \& Molau, S.\ 2015, WGN, J. IMO,  43, 43

\bibitem[Rudawska \& Jenniskens(2014)]{Rudawska}
Rudawska, R., \& Jenniskens, P.\ 2014, in Meteoroids 2013, T. J. Jopek et al. (eds.)
A. M. University Press, Pozna\'n, Poland, p. 217

\bibitem[Sekanina(1973)]{Sekanina1973}
Sekanina, Z.\ 1973, \icarus, 18, 253

\bibitem[Sekanina(1976)]{Sekanina1976}
Sekanina, Z.\ 1976, \icarus, 27, 265

\bibitem[Shiba(2017)]{Shiba2017}
Shiba, Y.\ 2017, WGN, J. IMO,  45, 127

\bibitem[Shiba(2022)]{Shiba2022}
Shiba, Y.\ 2022, WGN, J. IMO,  50, 38

\bibitem[SonotaCo(2009)]{SonotaCo}
SonotaCo 2009, WGN, J. IMO,  37, 55

\bibitem[Spurn{\'y} et al.(2017)]{Taurids}
Spurn\'y, P., Borovi\v{c}ka, J., Mucke, H., Svore\v{n}, J.  2017,
\aap, 605:A68

\bibitem[Trigo-Rodriguez et al.(2009)]{Trigo2009} 
Trigo-Rodriguez, J. M., Madiedo, J. M., Williams, I. P., \& Castro-Tirado, A. J.\ 2009, \mnras, 392, 367

\bibitem[Vaubaillon et al.(2019)]{Vaubaillon} 
Vaubaillon, J., Neslu{\v{s}}an, L., Sekhar, A.,  Rudawska R., \& Ryabova G.~O.\ 2019, 
in Meteoroids: Sources of Meteors on Earth and Beyond, 
G.~O. Ryabova, D.~J. Asher, \& M.~D. Campbell-Brown M. D. (eds.), 
Cambridge, UK: Cambridge University Press, 
p. 161

\bibitem[Whipple(1954)]{Whipple1954}
Whipple, F. L.\ 1954, \aj, 59, 201

\end{thebibliography}
\end{document}